\documentclass[a4paper,12pt]{article}

\usepackage{orcidlink,thumbpdf,lmodern}
\usepackage{amsmath}
\usepackage{amsthm}
\usepackage{amsfonts}
\usepackage[hypcap=false]{caption}
\usepackage{listings}
\usepackage{lscape}
\usepackage{natbib}
\usepackage{graphicx}
\usepackage{authblk}
\usepackage{geometry} 
\usepackage{graphicx}%
\usepackage{multirow}%
\usepackage{mathrsfs}%
\usepackage[title]{appendix}%
\usepackage{xcolor}%
\usepackage{textcomp}%
\usepackage{manyfoot}%
\usepackage{booktabs}%
\usepackage{algorithm}%
\usepackage{algorithmicx}%
\usepackage{algpseudocode}%
\usepackage[export]{adjustbox}
\usepackage{pdfpages}

\newtheorem{proposition}{Proposition}
\newtheorem{exa}{Example}%
\newtheorem{definition}{Definition}%
\newtheorem{lemma}{Lemma}

\newcommand{\vect}[1]{\ensuremath{\boldsymbol{#1}}}
\newcommand{\mat}[1]{\ensuremath{#1}}

\begin{document}

\title{A unified framework for multivariate two-sample and $k$-sample kernel-based quadratic distance goodness-of-fit tests}

\author[]{Marianthi Markatou\thanks{Corresponding author: \href{mailto:markatou@buffalo.edu}{markatou@buffalo.edu}}\orcidlink{0000-0002-1453-8229}}
\author[]{Giovanni Saraceno\thanks{Current affiliation: Dept. of Statistical Sciences, University of Padua \href{mailto:giovanni.saraceno@unipd.it}{giovanni.saraceno@unipd.it}}\orcidlink{0000-0002-1753-2367}}

\affil[]{Department of Biostatistics, University at Buffalo, Kimball Tower, Buffalo, NY, 14214, USA}

\date{}

\maketitle

\begin{abstract}

In the statistical literature, as well as in artificial intelligence and machine learning,  measures of discrepancy between two probability distributions are largely used to develop measures of goodness-of-fit.  
We concentrate on quadratic distances, which depend on a non-negative definite kernel.  
We propose a unified framework for the study of two-sample and $k$-sample goodness-of-fit tests based on the concept of matrix distance. We provide a succinct review of the goodness-of-fit literature related to the use of distance measures, and specifically to quadratic distances. We show that the quadratic distance kernel-based two-sample test has the same functional form with the maximum mean discrepancy test. We develop tests for the $k$-sample scenario, where the two-sample problem is a special case, derive their asymptotic distribution under the null hypothesis, and discuss computational aspects of the test procedures. We assess their performance, in terms of level and power, via extensive simulations and a real data example. The proposed framework is implemented in the \texttt{QuadratiK} package, available in both \textsf{R} and \textsf{Python} environments.

\smallskip 
    
\noindent \textbf{Keywords}:Goodness-of-fit, Kernel-based tests, $k$-sample test, Matrix distance, Tuning parameter selection.

\end{abstract}

\maketitle

The Goodness-of-Fit (GoF) problem has a long history in statistics, with a significant body of literature dedicated to univariate GoF testing. However, relatively less attention has been given to the two-sample and $k$-sample problem in the multivariate case. These GoF problems are particularly relevant in modern data analysis, where high-dimensional settings pose unique challenges.

Extensions of univariate two-sample GoF tests in high-dimensional, multivariate settings utilize graph-based multivariate ranking strategies. One such approach was introduced by \cite{Friedman1979}, who employed the minimal spanning tree (MST) as a multivariate ordered list. The authors extended the univariate run-based test introduced by \cite{Wald1940} and the univariate two-sample Kolmogorov-Smirnov test. Furthermore, the authors also proposed a modified version of the Kolmogorov-Smirnov test to address its relatively low power against scale alternatives. Similarly, \cite{Biswas2014a} adopted the shortest Hamiltonian path instead of MST as ranking strategy for multivariate observations, \cite{Chen2017} presented a test statistic based on a similarity graph. Another approach, proposed by \cite{Bickel1969} utilized Fisher's permutation principle to construct a multivariate two-sample Smirnov test, and \cite{Cao2006} introduced a local empirical likelihood test with a semiparametric extension for the multivariate two-sample GoF problem. 
In addition to graph-based and permutation-based methods, general distance measures have been widely adopted to deal with multivariate two-sample GoF testing, where the distance does not necessarily satisfy all the properties in the mathematical definition. Examples of such distance-based tests include \cite{Henze1988,Schilling1986,Mondal2015,Barakat1996,Chen2013}; \cite{Hall2002,Baringhaus2004,Szekely2004,Biswas2014b}; \cite{liu2011,Anderson1994,Fernandez2008}. Among these, \cite{Rosembaum2005} introduced a two-sample test based on crossmatch of inter-point distances. The use of distances in addressing the multivariate two-sample GoF problem has attracted attention not only from the field of statistics but also from various other disciplines. For example, \cite{Aslan2005}, from the field of physics, constructed a multivariate GoF test using a positive definite distance function to assess how accurately a model fits the observed data. In biology, \cite{Szabo2002} considered a negative quadratic distance to identify a subset of genes that differ the most. In the machine learning community, \cite{Gretton2012} proposed a test called Maximum Mean Discrepancy (MMD), which relies on the properties of the kernel mean embedding theory and it can be used with large samples and high-dimensional datasets. Furthermore, \cite{Mitchell2022} described how certain measures of statistical distance can be implemented as diagnostics for simulations involving charged-particle beams. We note that the operating characteristics of several of the tests mentioned above were studied in \cite{Chen2020}.

The multivariate $k$-sample goodness-of-fit problem extends the two-sample case to the scenario where there are more than two groups or samples to compare. In this setting, it is of interest to determine whether the distributions of the $k$ groups differ significantly from each other. To address this problem, various statistical techniques have been developed. For example, \cite{Kiefer1959,Wolf1973} and \cite{Scholz1987} extended the traditional Kolmogorov-Smirnov, Cram\'er-von Mises and Anderson-Darling tests, introduced for the case of two distributions, for treating the $k$-sample problem. \cite{Zhang2007} proposed a procedure based on likelihood ratio obtaining more powerful tests than the aforementioned approaches. \cite{Hukov2008} introduced multivariate tests based on the weighted $L_2$ distance between empirical characteristic functions; \cite{hlavka2021} adopted the same approach for constructing tests of serial independence for functional data, and \cite{Rizzo2016} proposed a multivariate $k$-sample test statistic based on energy statistics. 

Kernel-based methods have also gained popularity for the $k$-sample problem due to their ability to handle high-dimensional data. We note that \cite{balogoun2018} extended the kernel-based approach of \cite{harchaoui2008}, which utilizes the maximum Fisher discriminant ratio, to the multivariate $k$-sample problem. Furthermore, \cite{balogoun2022} proposed a $k$-sample test for functional data by introducing a generalization of the MMD.

In this article, our interest centers in developing novel GoF tests inspired by the kernel-based quadratic distances (KBQDs) introduced by \cite{Lindsay2008}. 
Quadratic distances are of interest for a variety of reasons including the following: (i) several important statistics such as Pearson's Chi-squared and Cram\'er-von Mises are exactly quadratic and many other distances are asymptotically quadratic - a valid von Mises expansion establishes their distributional equivalence to a quadratic distance; (ii) the empirical quadratic distance between two distributions has a relatively simple asymptotic theory; (iii) a simple interpretation of the distance as risk is possible; (iv) if minimum distance estimation is used, a simple analysis of the information potentially lost at the model is possible; (v) simple connections and interpretations in terms of robustness can be made (see \cite{Markatou2017}).
The kernel-based quadratic distance has been used to construct the multivariate one-sample goodness-of-fit tests by \cite{Lindsay2014}, and the asymptotic distribution under the composite null hypothesis and alternative hypothesis were also studied. 

Our contributions are as follows. We introduce a unified framework for the study of two-sample and $k$-sample tests that is based on the novel concept of \textit{matrix distance}. The elements of the matrix distance are used to derive two test statistics for testing equality of distributions of $k$ samples, and the two-sample statistic is derived as a special case of the $k$-sample statistic. We derive the asymptotic properties of the matrix distance and of the two test statistics under the null hypothesis when the kernel matrix has the form $\mat{K}(\vect{x}, \vect{y}) = (k^{\frac{1}{2}}(\vect{x}, \vect{y}) \vect{1})(k^{\frac{1}{2}}(\vect{x}, \vect{y}) \vect{1})^\top$, and address practical aspects related to the implementation of the test statistics. We further evaluate the performance of the $k$-sample and two-sample tests through a comprehensive simulation study. We demonstrate that the proposed tests exhibit high power against asymmetric alternatives that are close to the null hypothesis and with small sample size, as well as in the $k \ge 3$ sample comparison, for dimension $d>2$ and all sample sizes. To ensure ease of use, the proposed test statistics are readily available in the package \texttt{QuadratiK}, implemented both in \texttt{R} and Python. By presenting a unified framework within which the $k$- and two-sample testing can be discussed and a comprehensive analysis of relevant testing procedures, our aim is to contribute powerful tools and practical guidance for assessing the fit of distributions.

The rest of the paper is organized as follows. 
Section \ref{sec:kbqd} provides the main concepts related to the kernel-based quadratic distances. In Section \ref{sec:k-sample-test}  we introduce the kernel-based matrix distance along with the derived k-sample test statistics, and we investigate the asymptotic properties of the matrix distance and of the test statistics. Considerations on the implementation of the proposed test statistics are presented in Section \ref{sec:computation}. We evaluated the performance of the two-sample and $k$-sample tests through simulation studies and a real data application in Section \ref{sec:simulation}. Finally, concluding remarks are provided in Section \ref{sec:conclusion}. Relevant proofs can be found in the Appendix and additional proofs and simulation results can be found in the Supplementary Material.

\section{Kernel-based quadratic distances}
\label{sec:kbqd}

Consider two probability measures $F$ and $G$, and a non-negative definite symmetric kernel function $K$, possibly depending on $G$. The Kernel-Based Quadratic Distance (KBQD) between $F$ and $G$ is defined as 
\begin{equation}
\label{eqn:quadratic-distance}
    d_K(F,G) = \iint K(\vect{s},\vect{t}) d(F-G)(\vect{s}) d(F-G)(\vect{t}).
\end{equation}
KBQD tests, as indicated by the name, depend on kernels. 
We use a special type of kernel that is defined as follows. Let $\mathcal{S}$ be a sample space with measurable sets $\mathcal{B}$. Let $K(\vect{x},\vect{y})$ be a real-valued $\mathcal{B}$-measurable bounded positive definite kernel function on a measure space $(\mathcal{S},\mathcal{B}, M)$ such that
\begin{equation*}
\int_{\mathcal{S}} \int_{\mathcal{S}}   K^2(\vect{x},\vect{y}) dM(\vect{x})dM(\vect{y}) < \infty. 
\end{equation*}
This relationship will hold for many examples because $K$ is bounded and $M$ is a probability measure. Such a kernel is a Hilbert-Schmidt type.

An interesting family of kernel functions is the family of \textit{diffusion kernels} \citep{Ding2023}. Examples of diffusion kernels include the normal kernel and the Poisson kernel. These kernels are probability kernels that depend on a tuning parameter vector $\vect{h}$ and satisfy the diffusion equation given by
\begin{equation*}
	K_{h_1+h_2}(\vect{x},\vect{y}) = \int K_{h_1}(\vect{x},\vect{s})K_{h_2}(\vect{s},\vect{y}) d\vect{s},
\end{equation*}
where $\vect{h}_1$ and $\vect{h}_2$ are the associated kernel parameters.
\cite{Lindsay2008} pointed out that the kernel function generating a particular distance was not unique and introduced the model centered kernel. The $G$-centered kernel, $K_G$, of the kernel $K$ is defined as 
\begin{equation}
\label{eqn:Kcentering}
    K_G(\vect{s},\vect{t}) = K(\vect{s},\vect{t}) - K(\vect{s},G) - K(G,\vect{t}) + K(G,G), 
\end{equation}
where 
\begin{align*}
K(\vect{s},G) = & \int K(\vect{s},\vect{t})dG(\vect{t}), \quad K(G,\vect{t}) = \int K(\vect{s},\vect{t})dG(\vect{s}),\\
& \mbox{and} \quad K(G,G) = \iint K(\vect{s},\vect{t})dG(\vect{s})dG(\vect{t}).
\end{align*}
The kernel $K_G$ generates the same KBQD as the kernel $K$, and plays a crucial role in deriving the asymptotic distribution of the kernel-based multivariate GoF test statistics proposed in this article. The centered kernel-based quadratic distance has been used to construct the multivariate one-sample GoF tests by \cite{Lindsay2014}. In this context, let $G$ be a known null distribution. We wish to assess whether the distribution $F$ follows $G$, i.e. $H_0: F=G$. Centering the kernel with respect to $G$, the corresponding KBQD can be written as $ d(F,G) = \iint K_G(\vect{s},\vect{t}) dF(\vect{s}) dF(\vect{t})$.

\section{Kernel-based matrix distance and k-sample test \label{sec:k-sample-test}}

In this section, we introduce the concept of a matrix distance that provides a unification of the $k$-sample problem, where $k\ge 3$, with the two-sample problem.
Consider the $k$-sample GoF problem, where interest centers in testing the null hypothesis $H_0: F_1 = \cdots = F_k$, against the alternative $H_1: F_i \not = F_j$, for some $1 \le i \not = j \le k$. Consider random samples of i.i.d. observations $\vect{x}^{(i)}_{1},\vect{x}^{(i)}_{2},\ldots, \vect{x}^{(i)}_{n_i} \sim F_i$ and let $\hat{F}_i$ be the corresponding empirical distribution function. Assume that under the null hypothesis all the compared distributions are equal to some unknown distribution $\bar{F}$, that is, $F_1 = \cdots = F_k = \bar{F}$. Let $K(\cdot, \cdot)$ be a non-negative definite kernel function. We define the $k \times k$ matrix distance $\mat{D} = (D_{ij})$ with respect to the kernel $K$, as the matrix with the $ij$-th element given by
\begin{equation*}
	D_{ij} = \iint K(\vect{x},\vect{y}) d(F_i - \bar{F})(\vect{x})d(F_j - \bar{F})(\vect{y}), \qquad i,j \in \{1,\ldots,k\}.
\end{equation*}
By centering the kernel function with respect to the distribution $\bar{F}$, we have
\begin{equation}
\label{eqn:mat-element}
	D_{ij} = \iint K_{\bar{F}}(\vect{x},\vect{y}) dF_i(\vect{x})dF_j(\vect{y}).
\end{equation}
Employing the empirical distribution functions $\hat{F}_i$, $i \in \{1,\ldots,k\}$, the empirical version of the matrix distance $\hat{\mat{D}}_n$ has as off-diagonal elements the $V$-statistics  
\begin{equation*}
\label{eqn:mat-element-off}
	\hat{D}_{ij} = \frac{1}{n_i n_j} \sum_{\ell=1}^{n_i}\sum_{r=1}^{n_j}K_{\bar{F}}(\vect{x}^{(i)}_\ell,\vect{x}^{(j)}_r), \qquad \mbox{ for }i \not= j
\end{equation*}
and in the diagonal the $U$-statistics.
\begin{equation*}
\label{eqn:mat-element-diag}
	\hat{D}_{ii} = \frac{1}{n_i (n_i -1)} \sum_{\ell=1}^{n_i}\sum_{r\not= \ell}^{n_i}K_{\bar{F}}(\vect{x}^{(i)}_\ell,\vect{x}^{(i)}_r), \qquad \mbox{ for }i = j.
\end{equation*}

Computational considerations are important for kernel selection. For example, if a $d$-dimensional multivariate normal distribution is assumed under the null hypothesis, we would use the multivariate normal kernel and center it with a multivariate normal distribution with mean and variance matrix estimated from the sample. However, in practice the tested distributions $F_i$, $i \in \{1,\dots,k\}$, are usually unknown, as well as the true common distribution $\bar{F}$ is unknown. Then, we propose a non-parametric centering using the weighted average distribution
\begin{equation*}
    \bar{F} = \frac{n_1 \hat{F}_1 + \cdots + n_k \hat{F}_k}{n}, \quad \mbox{ with } n=\sum_{i=1}^k n_i.
\end{equation*}


The concept of matrix distance can be generalized as follows.
Consider an arbitrary vector of kernels $\vect{k}^\top(\vect{x},\vect{t}) = (k_1(\vect{x},\vect{t}), \ldots, k_d(\vect{x},\vect{t}))$. The construction 
\begin{equation}
\label{eqn:matrix-kernel}
    K(\vect{x},\vect{y}) = \int \vect{k}(\vect{x},\vect{t})\vect{k}^\top(\vect{y},\vect{t})du(\vect{t})
\end{equation}
creates a matrix kernel so that the corresponding matrix distance between $F$ and $G$ satisfies the usual definition of a quadratic distance. Before we formally define the concept of a matrix distance, we note here that the matrix kernel defined in equation (\ref{eqn:matrix-kernel}) is generally not symmetric. This matrix kernel can be symmetrized if we define
\begin{equation*}
    \mat{K}(\vect{x},\vect{y}) = \frac{1}{2} \int (\vect{k}(\vect{x},\vect{t}) \vect{k}^\top(\vect{y},\vect{t}) + \vect{k}(\vect{y},\vect{t}) \vect{k}^\top(\vect{x},\vect{t})) du(\vect{t}).
\end{equation*}
If the elements of the vector of kernels $\vect{k}^\top(\vect{x},\vect{t})$ are nonnegative definite kernels, then $\mat{K}(\vect{x},\vect{y})$ is a nonnegative definite kernel matrix. Indeed, considering the matrix kernel in equation (\ref{eqn:matrix-kernel}), for every $\vect{a} \in \mathbb{R}^d$
\begin{align*}
    \vect{a}^\top \mat{K}(\vect{x},\vect{y}) \vect{a} & = \vect{a}^\top \left(\int \vect{k}(\vect{x},\vect{t})\vect{k}^\top(\vect{y},\vect{t})du(\vect{t}) \right) \vect{a} \\
    & = \sum_{i=1}^d \sum_{j=1}^d a_i \left(\int k_i(\vect{x},\vect{t}) k_j^\top(\vect{y},\vect{t})du(\vect{t}) \right)a_j \ge 0,
\end{align*}
given that the product of nonnegative definite kernels is nonnegative definite. 
As a special case, if the elements of the kernel vector $\vect{k}(\vect{x},\vect{t})$ are root kernels \citep{Lindsay2014}, then the kernel $\mat{K}(\vect{x},\vect{y})$ generates a quadratic matrix distance with elements equal to the $D_{ij}$ defined above.

\begin{definition}
    Let $F,G$ be two probability distributions and $\mat{K}(\vect{x},\vect{y})$ a matrix kernel defined by (\ref{eqn:matrix-kernel}). Then, the matrix distance between $F,G$ is defined as 
    \begin{equation*}
    d_K(F,G) = \iint \mat{K}(\vect{x},\vect{y}) d(F-G)(\vect{x})d(F-G)(\vect{y}) \\
\end{equation*}
with $d_K(F,G) \ge 0$ in a matrix sense.
\end{definition}

The matrix distance $d_K(F,G)$ defined above is therefore a nonnegative matrix, and each of its elements is greater than or equal to 0, i.e. $\left(d_K(F,G)\right)_{ij} \ge 0$, $\forall i,j \in \{ 1, \ldots, d\}$.
Note that even if the kernel matrix $\mat{K}(\vect{x},\vect{y})$ defined in (\ref{eqn:matrix-kernel}) is not symmetric, the distance matrix $d_K(F,G)$ is symmetric. This is because of the following Proposition. 
\begin{proposition}
\label{prop:k-symmetric}
    Let $F,G$ be two probability distributions and $\mat{K}(\vect{x},\vect{y})$ the kernel matrix given in (\ref{eqn:matrix-kernel}), with $\vect{k}^\top(\vect{x},\vect{t})$ being a vector of nonnegative definite kernels. Then, the matrix distance $d_K(F,G)$ is symmetric. 
\end{proposition}
\begin{proof}
    The proof is reported in Section S1 of the Supplementary Material.
\end{proof}

Note that we do not require $\vect{k}(\vect{x},\vect{t})$ to be a vector of symmetric kernels for $d_K(F,G)$ to be symmetric. 

In what follows, we present an example that illustrates the construction of the matrix distance with elements defined in equation (\ref{eqn:mat-element}) for the comparison of three samples. 

\begin{exa}[$\mathbf{k=3}$]
    Consider the case $k=3$, and the null hypothesis $H_0: F_1 = F_2 = F_3$ against the alternative $H_1: F_i\not=F_j$ for $i,j \in \{1,2,3\}$. Let $\vect{x}_1^{(i)}, \ldots, \vect{x}_{n_1}^{(i)} \sim F_i$ be a i.i.d. sample for $i \in \{1,2,3\}$ and $\bar{F} = (n_1F_1 + n_2F_2 + n_3F_3)/n$ with $n=n_1+n_2+n_3$. Then, the introduced matrix distance has the form
    \begin{equation*}
        \mat{D} =
    \begin{pmatrix}
    D_{11} & D_{12} & D_{13} \\
    D_{21} & D_{22} & D_{23} \\
    D_{31} & D_{32} & D_{33} \\
    \end{pmatrix},
    \end{equation*}
    that is
    \begin{equation*}
    \begin{pmatrix}
    \iint K_{\bar{F}}(\vect{x},\vect{y}) \, dF_1(\vect{x}) \, dF_1(\vect{y}) & \iint K_{\bar{F}}(\vect{x},\vect{y}) \, dF_1(\vect{x}) \, dF_2(\vect{y}) & \iint K_{\bar{F}}(\vect{x},\vect{y}) \, dF_1(\vect{x}) \, dF_3(\vect{y}) \\
    \iint K_{\bar{F}}(\vect{x},\vect{y}) \, dF_2(\vect{x}) \, dF_1(\vect{y}) & \iint K_{\bar{F}}(\vect{x},\vect{y}) \, dF_2(\vect{x}) \, dF_2(\vect{y}) & \iint K_{\bar{F}}(\vect{x},\vect{y}) \, dF_2(\vect{x}) \, dF_3(\vect{y}) \\
    \iint K_{\bar{F}}(\vect{x},\vect{y}) \, dF_3(\vect{x}) \, dF_1(\vect{y}) & \iint K_{\bar{F}}(\vect{x},\vect{y}) \, dF_3(\vect{x}) \, dF_2(\vect{y}) & \iint K_{\bar{F}}(\vect{x},\vect{y}) \, dF_3(\vect{x}) \, dF_3(\vect{y})
    \end{pmatrix},
    \end{equation*}
    where $K_{\bar{F}}$ denotes the kernel centered with respect to $\bar{F}$. Notice that, the elements in the diagonal correspond to the kernel-based quadratic distances between each $F_i$ and $\bar{F}$, while the off diagonal elements measure the contrasts for each pair of the tested distributions. 
    The kernel function is centered with respect to the weighted average distribution $\bar{F} = (n_1F_1 + n_2F_2 + n_3F_3)/(n_1 + n_2 + n_3)$ and the matrix distance $\mat{D}$ is orthogonal to the vector $(n_1/n, n_2/n, n_3/n)^\top$. Alternatively, an equally weighted $\bar{F}$ could be used, which would then create a matrix orthogonal to $(1,1,1)^\top$. In this case, the eigenvectors are contrasts, aiding the interpretation of the results. 
    
    An eigenanalysis of the matrix distance provides the following information. The leading eigenvalue of $\mat{D}$ gives the linear combination $a_1(F_1 - \bar{F}) + a_2(F_2 - \bar{F}) + a_3(F_3 - \bar{F})$ that is most different from zero. The null space of the matrix $\mat{D}$ consists of linear combinations $\sum_{i=1}^3 b_i(F_i - \bar{F}) = 0$. For example, if $F_3 - \bar{F} = 0$, which occurs when $F_3 = n_1(n_1 +n_2)^{-1}F_1 + n_2(n_1 +n_2)^{-1}F_2$, then the vector $(1, -1, 0)$ is in the null space of $\mat{D}$.
\end{exa}

We now investigate the asymptotic distribution of the matrix distance $\hat{\mat{D}}_{n}$ under the null hypothesis. Notice that, each element of this matrix distance depends on the same symmetric root kernel $k^{\frac{1}{2}}(\vect{x},\vect{t})$, hence the $ij$-element of the matrix distance $\mat{K(\vect{x},\vect{y})}$ is given as $\int k^{\frac{1}{2}}(\vect{x},\vect{t})k^{\frac{1}{2}}(\vect{t},\vect{y}) du(\vect{t})$. See \cite{Lindsay2014} for a discussion on the role of root kernels in this context. The first result is given by the spectral decomposition of the centered kernel under $\bar{F}$. 

\begin{proposition}
\label{prop:Dij-spectral}
	Let $\mat{D}$ be the matrix distance with elements defined in equation (\ref{eqn:mat-element}). The matrix distance $\mat{D}$ can be written as 
\begin{equation*}
	\mat{D} = \sum_{p=1}^\infty \lambda_p \tilde{\vect{\phi}}_{p} \tilde{\vect{\phi}}_{p}^\top ,
\end{equation*}
where $\tilde{\vect{\phi}}_{p} = (\bar{\phi}_{p,1},\ldots,\bar{\phi}_{p,k})$ is the $k$-dimensional vector with entries, given as    
\begin{equation*}
	\bar{\phi}_{p,i} = \int \phi_p(\vect{x}) dF_i(\vect{x}),
\end{equation*}
are averages of mean-zero, variance-1 random variables uncorrelated with respect to $p$. 
\end{proposition}
\begin{proof}
    The proof is reported in Section S2 of the Supplementary Material.
\end{proof}

Consider random samples of i.i.d. observations $\vect{x}^{(i)}_{1},\vect{x}^{(i)}_{2},\ldots, \vect{x}^{(i)}_{n_i} \sim F_i$, $i \in \{1, \ldots, k\}$. The asymptotic distribution of the matrix distance $\hat{\mat{D}}_{n}$ under the null hypothesis is provided as follows. 
\begin{proposition}\label{prop:asym-D}
Let $K_{\bar{F}}$ be the kernel centered by the common distribution $\bar{F}$ such that
\begin{itemize}
    \item[](i) $ \int K_{\bar{F}}(\vect{t},\vect{t}) d\bar{F}(\vect{t}) < \infty$ ;
    \item[](ii) $\iint K_{\bar{F}}^2(\vect{s},\vect{t}) d\bar{F}(\vect{s})d\bar{F}(\vect{t}) < \infty$;
\end{itemize}
Then, under the null hypothesis $H_0:F_1=\cdots=F_k$ we have that
\begin{equation}
\label{eqn:asym-distr-k}
    n\hat{\mat{D}}_{n}  \stackrel{d}{\longrightarrow} W^\ast = \sum_{p=1}^\infty \lambda_p W_{k,p},
\end{equation}
where $\lambda_p$ are the nonzero eigenvalues of $K_{\bar{F}}$; $W_{k,p}$, $p=1,2,\ldots$, are independent random variables that follow the $k$-dimensional Wishart distribution with covariance matrix $\mat{V}$ and $1$ degree of freedom, denoted as $W_{k,p} \sim \mathcal{W}_k(1, \mat{V})$, and $\mat{V}$ is a diagonal matrix with $V_{ii}=1/\rho_i$ with $\rho_i = \lim_{n,n_i \to \infty} n_i/n$ and $0 < \rho_i < 1$.
\end{proposition}
\begin{proof}
    The proof is reported in Section \ref{app:prop3} of the Appendix.
\end{proof}

\begin{exa}[\textbf{k=3 -- Continued}]
   For $k=3$, when the kernel is centered with respect to the weighted average distribution $\bar{F}$ the covariance matrix of the independent Wishart distribution is given as $\mat{V} = \mathrm{diag}\left(n/n_1,n/n_2,n/n_3\right)$.

    Notice that, if the non-centered kernel is employed the spectral decomposition of the matrix distance $\mat{D}$ is given as
    \begin{equation*}
        \mat{D}  = \sum_{p=1}^\infty \lambda_p \vect{\phi}^\ast_{p} (\vect{\phi}^\ast_{p})^\top 
    \end{equation*}
    with $\vect{\phi}^\ast_{p} = (\bar{\phi}_{p,1} - \bar{\phi}_{p},\ldots,\bar{\phi}_{p,k} - \bar{\phi}_{p})$, where
    \begin{equation*}
    	\bar{\phi}_{p} = \int \phi_p(\vect{x}) d\bar{F}(\vect{x}).
    \end{equation*}
    Let $\vect{\pi}^\top = (n_1/n, \ldots, n_k/n)$ and let $\vect{1}$ denote a vector of ones. then, $\vect{\phi}^\ast_{p}$ can be written as $\vect{\phi}^\ast_{p} = \mat{I}^\ast \tilde{\vect{\phi}}_{p}$ with $\mat{I}^\ast = I - \vect{1}\vect{\pi}^\top$, and in this case the variance of the Wishart distribution is given as $\mat{V}^\ast = \mat{I}^\ast\mat{V}(\mat{I}^\ast)^\top$. For $k=3$
    \begin{equation*}
    \mat{V}^\ast =
    \begin{pmatrix}
    \frac{n_{(1)}}{n_1} & -1 & -1 \\
   -1 & \frac{n_{(2)}}{n_2} & -1 \\
    -1 & -1 & \frac{n_{(3)}}{n_3} \\
    \end{pmatrix},
    \end{equation*}
    where $n_{(i)}$ denotes the sum of the samples sizes except for $n_i$. Note that, the matrix $\mat{V}^\ast$ is not full rank.  
    
\end{exa}

\subsection{\textit{k}-Sample Tests}

In order to perform hypothesis testing in the $k$-sample problem against general alternatives $H_1: F_i \not= F_j$, for some $1 \leq i\not=j \leq k$, we propose two tests derived from the matrix distance. Specifically, we consider the trace statistic
\begin{equation}
\label{eqn:k-sample-stat-trace}
	\mathrm{trace}(\hat{\mat{D}}_n) =  \sum_{i=1}^{k}\hat{D}_{ii}.
\end{equation}
and $T_n$, derived considering all the possible pairwise comparisons in the $k$-sample null hypothesis, given as
\begin{equation}
\label{eqn:k-sample-stat}
	T_{n} = (k-1) \mathrm{trace}(\hat{\mat{D}}_n) - 2 \sum_{i=1}^{k}\sum_{j> i}^{k}\hat{D}_{ij}.
\end{equation}

In what follows, we discuss the asymptotic distribution of the proposed test statistics. We begin with the trace statistic.

\begin{proposition}
\label{prop:asym-trace}
    Let $n=n_1+\cdots + n_k$ and let $K_{\bar{F}}$ be the kernel centered by a common distribution $\bar{F}$ such that
\begin{itemize}
    \item[](i) $ \int K_{\bar{F}}(\vect{t},\vect{t}) d\bar{F}(\vect{t})< \infty$ ;
    \item[](ii) $\iint K_{\bar{F}}^2(\vect{s},\vect{t}) d\bar{F}(\vect{s})d\bar{F}(\vect{t}) < \infty$;
    \item[](iii) $K_{\bar{F}}$ is continuous at $(\vect{s},\vect{s})$ for almost all $\vect{s}$ with respect to the measure $\bar{F}$.
\end{itemize}
Then, under the null hypothesis $H_0:F_1=\cdots=F_k=\bar{F}$ 
\begin{equation}
\label{eqn:asym-distr-trace}
    n\mathrm{trace}(\hat{D}_{n})  \stackrel{d}{\longrightarrow}  \sum_{p=1}^\infty \lambda_p \sum_{i=1}^k \frac{1}{\rho_i} \left( Z_{ip}^2- 1\right) \mbox{ as } n_i,n \to \infty,
\end{equation}
where $Z_{1p},\ldots,Z_{kp}$ are $k$ independent random variables that follow the standard normal distribution; $\lambda_p$ are the nonzero eigenvalues of $K_{\bar{F}}$ and $\rho_i= \lim_{n_i,n\to\infty} n_i/n$, $i \in \{1, \ldots, k\}$ and $0<\rho_i<1$.
\end{proposition}
\begin{proof}
    See Section \ref{app:asym-trace} of the Appendix. 
\end{proof}

The asymptotic distribution of $T_{n}$ under the null hypothesis is provided in the following Proposition. 
\begin{proposition}
\label{prop:asym-norm-test}
Let $n=n_1+\cdots + n_k$ and let $K_{\bar{F}}$ be the kernel centered by the common distribution $\bar{F}$ such that
\begin{itemize}
    \item[](i) $ \int K_{\bar{F}}(\vect{t},\vect{t}) d\bar{F}(\vect{t})< \infty$ ;
    \item[](ii) $\iint K_{\bar{F}}^2(\vect{s},\vect{t}) d\bar{F}(\vect{s})d\bar{F}(\vect{t}) < \infty$;
    \item[](iii) $K_{\bar{F}}$ is continuous at $(\vect{s},\vect{s})$ for almost all $\vect{s}$ with respect to the measure $\bar{F}$.
\end{itemize}
Then, under the null hypothesis $H_0:F_1=\cdots=F_k=\bar{F}$ we have that
\begin{equation}
\label{eqn:asym-distr}
    nT_{n}  \stackrel{d}{\longrightarrow} \chi^\ast = \sum_{p=1}^\infty \lambda_p \left[ \sum_{i=1}^k \sum_{j>i}^k\left( \frac{1}{\sqrt{\rho_i}}Z_{ip}- \frac{1}{\sqrt{\rho_j}}Z_{jp}\right)^2 - \sum_{i=1}^k \frac{1}{\rho_i} \right] \mbox{ as } n_i,n \to \infty,
\end{equation}
where $Z_{1p},\ldots,Z_{kp}$ are $k$ independent random variables that follow the standard normal distribution; $\lambda_p$ are the nonzero eigenvalues of $K_{\bar{F}}$; $\rho_i= \lim_{n_i,n\to\infty} n_i/n$, $i \in \{1, \ldots, k\}$ and $0<\rho_i<1$.
\end{proposition}
\begin{proof}
    The proof is reported in Section \ref{app:prop5} of the Appendix.
\end{proof}

\subsection{Two-sample goodness-of-fit Tests}
\label{subsec:two-sample}

As highlighted in the Introduction, the problem of testing the general non parametric hypothesis $H_0: F=G$ versus the alternative $H_1:F\not=G$ with $F$ and $G$ unknown, is of particular interest. Assume that $\vect{x}_1, \ldots, \vect{x}_n$ is a random sample from the distribution of the random variable $\vect{X} \sim F$ and $\vect{y}_1, \ldots, \vect{y}_m$ is a random sample from the distribution of the random variable $\vect{Y} \sim G$.
The KBQD test statistics in equations (\ref{eqn:k-sample-stat-trace}) and (\ref{eqn:k-sample-stat}) reduce to  
\begin{equation*}
    \mathrm{trace}(\hat{\mat{D}}_n) =  \frac{1}{n(n-1)}\sum_{i=1}^n \sum_{j \not=i}^n K_{\bar{F}}(\vect{x}_i,\vect{x}_j) + \frac{1}{m(m-1)}\sum_{i=1}^m \sum_{j \not=i}^m K_{\bar{F}}(\vect{y}_i,\vect{y}_j),
\end{equation*}
and
\begin{align}
    \label{eqn:two-sample-statistic}
    T_{n,m} =  \frac{1}{n(n-1)}&\sum_{i=1}^n \sum_{j \not=i}^n K_{\bar{F}}(\vect{x}_i,\vect{x}_j) - \frac{2}{nm}\sum_{i=1}^n \sum_{j =1}^m K_{\bar{F}}(\vect{x}_i,\vect{y}_j) \\
    & + \frac{1}{m(m-1)}\sum_{i=1}^m \sum_{j \not=i}^m K_{\bar{F}}(\vect{y}_i,\vect{y}_j).  \nonumber
\end{align}
Notice that, the kernel function is centered with respect to 
\begin{equation*}
    \bar{F} = \frac{n}{n+m}\hat{F} + \frac{m}{n+m}\hat{G}.
\end{equation*}
\begin{lemma}\label{lemma:1}
    For any centering distribution $F^\ast$
    \begin{equation*}
        d_K(F,G) = d_{K_{F^\ast}}(F,G),
    \end{equation*}
    where $d_K(F,G)$ is defined in equation (\ref{eqn:quadratic-distance}).
\end{lemma}
\begin{proof}
    See Section S3 of the Supplementary Material.
\end{proof}
The asymptotic distribution of the two-sample KBQD test is the same as that of the MMD test and it can be found in \citet[Appendix B1]{Gretton2012}. 
The case $k=2$ is considered in Section S4 of the Supplementary Material. 

\subsection{Kernel-based tests in the literature}
\label{subsec:kb-test-literature}

At this point, it is instructive to visit the constructions of kernel tests that exist in the literature.
\cite{rao1982} introduced the concept of an exact quadratic distance for discrete population distributions named \textit{quadratic entropy}. Rao's definition of quadratic entropy is as follows: 
\begin{equation*}
    Q(P) = \iint K(\vect{x},\vect{y}) dP(\vect{x}) dP(\vect{y}),
\end{equation*}
where $K(\vect{x},\vect{y})$ is a measurable, conditionally negative definite kernel, i.e. the kernel satisfies the condition 
\begin{equation*}
    \sum_{i=1}^n\sum_{j=1}^n K(\vect{x}_i,\vect{x}_j) a_i a_j \le 0, 
\end{equation*}
for all $\vect{x}_1,\ldots,\vect{x}_n$ that belong to the sample space and any choice of real numbers $a_1, \ldots, a_n$ such that $\sum_{i=1}^n a_i = 0$. Quadratic entropy has several interesting properties, such as the convexity of Jensen differences of all orders (see \cite{rao1982}). This is the concept closest to the work presented here. Other relevant works are as follows.

Let $\vect{X}$ and $\vect{Y}$ be random variables on the sample space $\mathcal{X}$ with corresponding probability measures $F$ and $G$. Given $\mathcal{F}$, a class of functions $f:\mathcal{X} \to \mathbb{R}$, the maximum mean discrepancy is defined as 
\begin{equation*}
    \text{MMD}[\mathcal{F},F,G] := \sup_{f \in \mathcal{F}} (E_{\vect{X}}[f(X)] - E_{\vect{Y}}[f(Y)]).
\end{equation*}
\cite{Gretton2012} consider as $\mathcal{F}$ the unit ball in a reproducing kernel Hilbert space (RKHS) $\mathcal{H}$, which has an associated kernel function $K$. The kernel mean embedding of the random variables $\vect{X}$ and $\vect{Y}$ are given as $\mu_{\vect{X}} = E_{\vect{X}[K(\vect{X}, \cdot)]}$ and $\mu_{\vect{Y}} = E_{\vect{Y}[K(\cdot, \vect{Y})]}$.
Within this setting, the \textbf{squared} MMD has the form
\begin{equation*}
    \text{MMD}^2(F,G) := ||\mu_{\vect{X}} - \mu_{\vect{Y}}||^2_{\mathcal{H}},
\end{equation*}
where $||\cdot||_{\mathcal{H}}$ denotes the Hilbert space norm. Then, the corresponding $U$-statistic is given by 
\begin{align*}
    \text{MMD}^2(\vect{X},\vect{Y}) =  \frac{1}{n(n-1)}&\sum_{i=1}^n \sum_{j \not=i}^n K(x_i,x_j) - \frac{2}{nm}\sum_{i=1}^n \sum_{j =1}^m K(x_i,y_j) \\
    & + \frac{1}{m(m-1)}\sum_{i=1}^m \sum_{j \not=i}^m K(y_i,y_j).  
\end{align*}
The centering of the kernel plays a crucial role in the computation of the asymptotic distribution of the kernel-based test statistics.
Furthermore, the selection of the kernel tuning parameter, as proposed by \cite{Gretton2012}, is determined by the median Euclidean distance within the pooled sample. This pragmatic approach offers a direct method for parameter estimation, however, it is acknowledged that this choice does not come with a formal optimality guarantee. 

\noindent
Another powerful method for comparing statistical distributions is given by the energy distance, defined as 
\begin{equation*}
\mathcal{E}(X, Y) = 2E[||X - Y||] - E[||X - X'||] - E[||Y - Y'||],
\end{equation*}
where $X'$ and $Y'$ are independent and identically distributed copies of $X$ and $Y$, respectively. The energy distance is a measure of the statistical distance between the distributions of the random vectors $\vect{X}$ and $\vect{Y}$. The corresponding statistic for the energy distance is
\begin{align*}
\mathcal{E}_{n,m}(X, Y) = \frac{2}{nm} \sum_{i=1}^n \sum_{j=1}^m ||x_i - y_j|| - \frac{1}{n^2} \sum_{i=1}^n \sum_{j=1}^n ||x_i - x_j|| - \frac{1}{m^2} \sum_{i=1}^m \sum_{j=1}^m ||y_i - y_j||.
\end{align*}
The presented literature belongs to the class of methods known as integral probability metrics, which measure the distance between probability distributions based on expectations of functions belonging to a certain class. Considering the MMD and energy distances, the primary difference between these two metrics lies in the function space $\mathcal{F}$: MMD utilizes functions from an RKHS defined by a kernel, whereas energy distance uses Euclidean distance-based functions. This difference in the function space leads to different properties and performance in statistical tests. For example, the energy distance has connections to the Cramér-von Mises criterion and is known for its strong performance in high-dimensional settings \citep{SZEKELY2013, Rizzo2016}.
While MMD and energy distance both provide robust methods for comparing distributions, their formulations and underlying assumptions differ. MMD leverages the rich structure of RKHS, offering flexibility through kernel choice and tuning, whereas energy distance directly utilizes the geometric properties of the sample space. 
The KBQD can encompass the presented distances through an appropriate choice of the kernel function. For example, the energy distance can be obtained by considering the linear kernel. \cite{Lindsay2014} offered a second interpretation of the quadratic distance as the $L_2$ distance between the smoothed version of two probability distributions by using the root kernel, indicating the connection of the KBQD with the reproducing kernel Hilbert space formulation employed in the MMD construction. 

\noindent
For the $k$-sample problem, with null hypothesis $H_0: F_1 = F_2 = \cdots = F_k$, \cite{balogoun2022} introduced the generalized maximum mean discrepancy (GMMD) as
\begin{equation*}
\text{GMMD}^2(F_1, \ldots, F_k, \vect{\pi}) = \sum_{i=1}^k \sum_{j\neq i}^k \pi_j \text{MMD}^2(F_i, F_j),
\end{equation*}
with $\vect{\pi} = (\pi_1, \ldots, \pi_k) \in (0,1)^k$ and $\sum_{i=1}^k \pi_i = 1$. \cite{balogoun2022} considered $\pi_i = n_i/n$. Notice that the GMMD statistic coincides with the proposed $T_n$ statistic for $\pi_i = 1$. 
As a final note, we remark that the connection between kernel-based and distance-based methods has been studied in the literature. The interested reader can consult \cite{sejdinovic2013} and \cite{sen2021}.

\section{Numerical aspects}
\label{sec:computation}

The kernels used in the relationship (\ref{eqn:mat-element}) belong to the class of diffusion kernels defined in \cite{Ding2023}. In this work, we use the normal kernel defined as 
\begin{equation*}
    K(\vect{x}, \vect{y}) = \frac{1}{(2\pi)^{d/2}|\mat{\Sigma}|^{1/2}} \exp\left\{-\frac{(\vect{x} - \vect{y})^\top \mat{\Sigma}^{-1} (\vect{x} - \vect{y})}{2}\right\}
\end{equation*}
where $\mat{\Sigma}$ is a matrix of tuning parameters. In the present paper, to simplify the computation, we specify $\mat{\Sigma} = h^2 \mat{I}$, so the kernel has the simpler form
\begin{equation*}
    K(\vect{x}, \vect{y}) = \frac{1}{(2\pi)^{d/2}h^d} \exp\left\{-\frac{(\vect{x} - \vect{y})^\top (\vect{x} - \vect{y})}{2h^2}\right\}.
\end{equation*}
Notice that this kernel uses the same smoothing parameter $h^2$ for all dimensions of the data vector. An alternative smoothing matrix is given by $\mat{\Sigma} = \mathrm{diag}(h_1^2, h_2^2, \ldots, h_d^2)$ with much higher computational complexity. \cite{Lindsay2014} discuss extensively the selection of the tuning parameter in the context of one-sample goodness-of-fit testing. In this work, we extend their algorithm in the case of $k$-sample testing.

In order to avoid any assumption on the common distribution $\bar{F}$ under the null hypothesis, we consider non-parametric centering with $\bar{F}=(n_1 F_1 + \cdots + n_k F_k)/n$ where $n=\sum_{i=1}^k n_i$. 
Let $\vect{z}_1,\ldots, \vect{z}_n$ denote the pooled sample. Following equation (\ref{eqn:Kcentering}), for any $s,t \in \{\vect{z}_1,\ldots, \vect{z}_n\}$, the non-parametric $\bar{F}$-centered kernel is given by 
\begin{align*}
    K_{\bar{F}}(\vect{s},\vect{t}) = &
    K(\vect{s},\vect{t}) - \frac{1}{n}\sum_{i=1}^{n} K(\vect{s},\vect{z}_i) - \frac{1}{n}\sum_{i=1}^{n} K(\vect{z}_i,\vect{t}) + \frac{1}{n(n-1)}\sum_{i=1}^{n} \sum_{j \not=i}^{n} K(\vect{z}_i,\vect{z}_j).  
\end{align*}
\cite{Chen2020} detailed both parametric and non-parametric centering of the test statistic for the two-sample problem. Both types of centering are implemented in the package \texttt{QuadratiK}, in \textsf{R} and \textsf{Python} \citep{saraceno2024goodnessoffit}, as available options when performing the two-sample test.   

An aspect closely related to the implementation is the calculation of the critical value. The asymptotic distributions of the test statistics involve the eigenvalues of the centered kernel. This step, while theoretically significant, is practically unnecessary. By employing numerical techniques such as the bootstrap, permutation and subsampling algorithms, we can compute the empirical critical value efficiently without the need for eigenvalue analysis. These methods not only simplify the process, but also ensure that the calculation is rigorous and applicable to real-world scenarios. The consistency of these resampling methods ensures that the tests achieve the nominal type I error probability. For example \cite{gonzalez2024} show the bootstrap consistency for the energy statistic, and \cite{Lee2021} analyze bootstrap calibration of the KBQD tests in goodness-of-fit problems. In the following, we provide numerical simulations to illustrate that the resampling-based procedures provide valid approximations under general conditions. The general algorithm for computing the critical value is illustrated in Algorithm \ref{alg:cv}.

\begin{algorithm}[H]
\caption{Algorithm for the computation of the critical value.}\label{alg:cv}
\begin{algorithmic}[1]
\State Let $B$ denote the number of sampling repetitions.
\State Generate k-tuples, of total size $n_B$, from the pooled sample $\vect{z}_1,\ldots, \vect{z}_n$ following one of sampling method (bootstrap/permutation/subsampling);
\State Compute the k-sample test statistic;
\State Repeat steps \texttt{1}-\texttt{2} $B$ times;
\State Select the 95$^{th}$ quantile of the obtained list of values. 
\end{algorithmic}
\end{algorithm}
The main difference between the bootstrap and subsampling algorithms is that they correspond to sampling from the pooled sample with and without replacement, respectively. While in bootstrap the samples are generated using the same sample size, the subsampling samples have smaller sample size than the original data, i.e. $n_B = b * n$ and $b \in (0,1]$. The determination of the ``optimal" sample size $n_B$ remains a subject without definitive guidelines.
Approaches for obtaining the optimal subsamples explored in the literature involve concepts of optimal subsampling probabilities. For details see \cite{yao_wang_2021}. We investigated the performance of Algorithm \ref{alg:cv} with the subsampling algorithm for different choices of the subsample proportion $b$, through a small simulation study reported in Section S5 of the Supplementary Material. In the simulation studies considered here we set $b=0.8$.
Finally, the permutation algorithm draws observations from the original data set, creating new samples that maintain the original sample size. This method aligns with the subsampling algorithm with $b=1$, indicating that new samples are formed by sampling without replacement. 

\begin{figure}[htb]
    \centering
    \includegraphics[width=\textwidth]{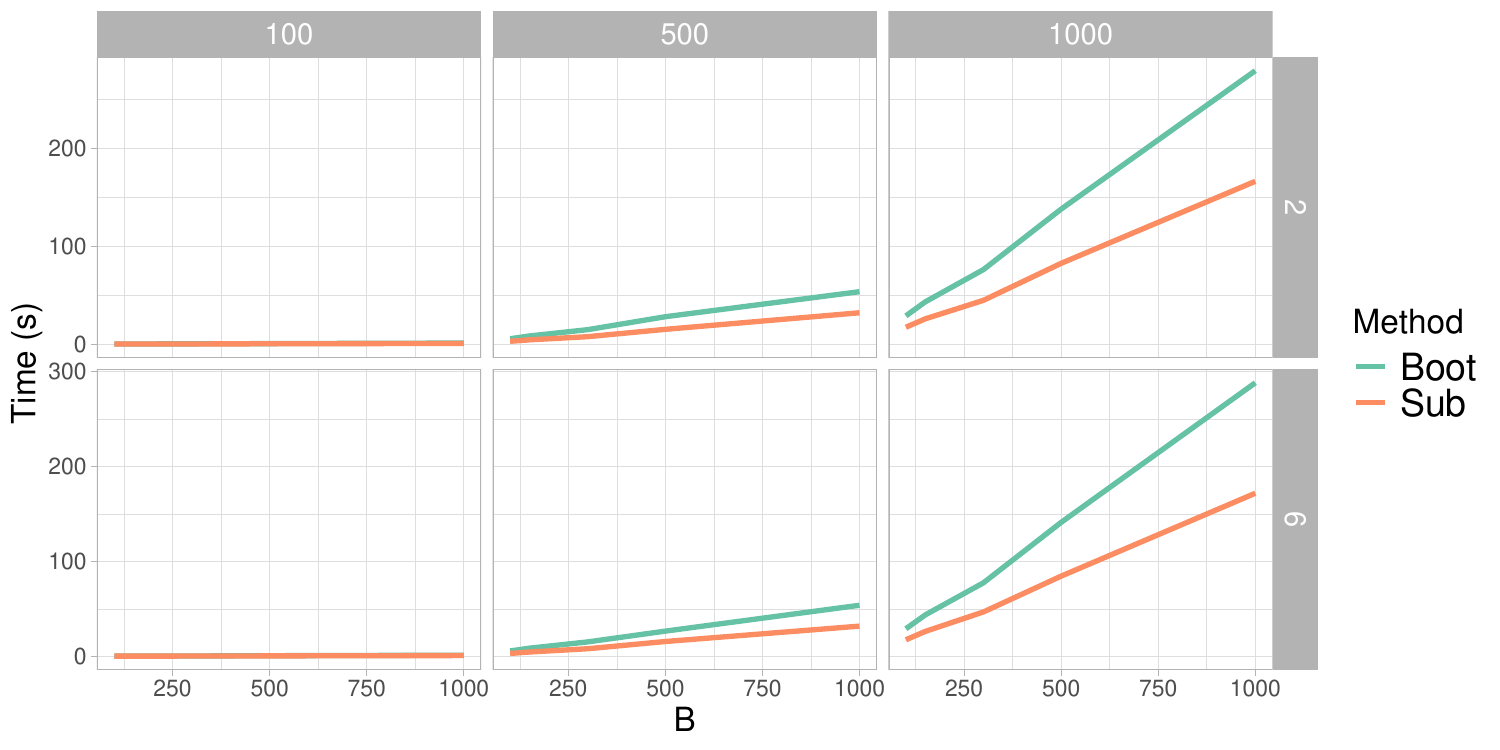}
    \caption{Average computational time (seconds) for computing the KBQD test for increasing number of replications $B$ for the bootstrap and  subsampling algorithms. One sample is generated from the $d$-dimensional standard normal distribution, while the second sample is generated from the skew-normal distribution $SN_d(\vect{0},\mat{I}_d,\vect{\lambda})$, with $\vect{\lambda} = \lambda \vect{1}$ and $\lambda = 0.1$. Similar trends were observed for $\lambda = 0, 0.2, 0.3$. The sample size $n=100, 500, 1000$ and dimension $d=2,6$, are indicated as headers. For the subsampling algorithm we use $b=0.8$. The permutation method exhibits the same computational time as bootstrap, and it is reported in Section S6 of the Supplementary Material.}
    \label{fig:time-B}
\end{figure}
The non-parametric calculation of the critical value depends on the number of replications of the bootstrap-, permutation- or subsampling-step, denoted as $B$. We investigate the performance of the proposed KBQD tests with respect to different values of $B$, when $k=2$. In particular, we consider $B= 100, 150, 300, 500, 1000$. We generated samples $\vect{x}_1,\ldots,\vect{x}_{n_1}$ from the normal distribution $N_d(\vect{0}, \mat{I}_d)$ and $\vect{y}_1,\ldots, \vect{y}_{n_2}$ from the skew-normal distribution $SN_d(\vect{0},\mat{I}_d,\vect{\lambda})$, where $\mat{I}_d$ denotes the $d$-dimensional identity matrix, $\vect{\lambda} = \lambda \vect{1}$ and $\lambda = 0, 0.1, 0.2, 0.3$. Observations are sampled using the \texttt{R} package ``\texttt{sn}". We considered dimensions $d=1,2,6$, sample size $n_1=n_2=100,500,1000$ and $N=1000$ replications. The tuning parameter $h$ of the KBQD test takes values from 0.2 to 10 with a step of 0.4, denoted as $h = 0.2(0.4)10$. 
Figure~\ref{fig:time-B} shows the mean computational time needed for computing the KBQD tests when $B$ increases, with respect to the considered algorithms, dimension $d=2,6$ and the sample size $n$ are indicated as header. The mean values are summarized with respect to the values of $h$, $\lambda$ and $N$ for a total of 100000 observations. The computational time increases linearly with increasing $B$. The subsampling algorithm requires about half the time needed by bootstrap and permutation procedures. This becomes relevant for very large sample sizes. Additionally, the computational time does not increase for higher dimensions. The permutation method exhibits results similar to the bootstrap sampling, and it is not reported here. Complete results are available in Section S6 of the Supplementary Material. 
\begin{figure}[htb]
    \centering
    \includegraphics[width=\textwidth]{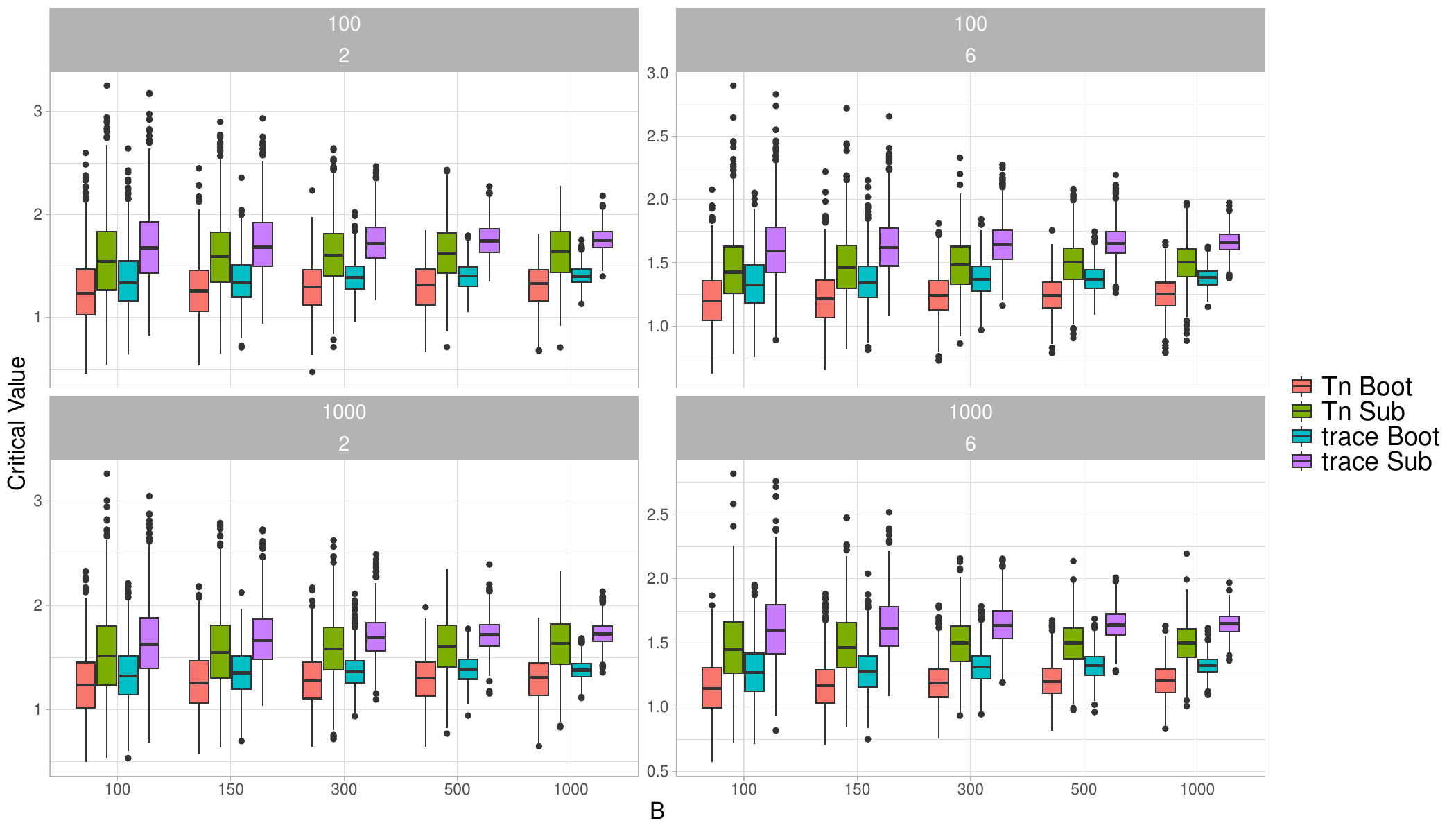}
    \caption{Boxplots of critical values of the KBQD test statistics, $T_n$ and $\mathrm{trace}$, with $h=2.2$, for increasing $B$ with respect to the bootstrap and subsampling algorithms. The two samples are generated from the $d$-dimensional standard normal distribution. The dimension $d$ and sample size $n$ are indicated as headers. The boxplots for the permutation sampling are similar to the bootstrap, and are reported in Section S6 of the Supplementary Material.}
    \label{fig:boxplot-B}
\end{figure}
\begin{figure}[!t]
    \centering
    \includegraphics[width=0.95\textwidth]{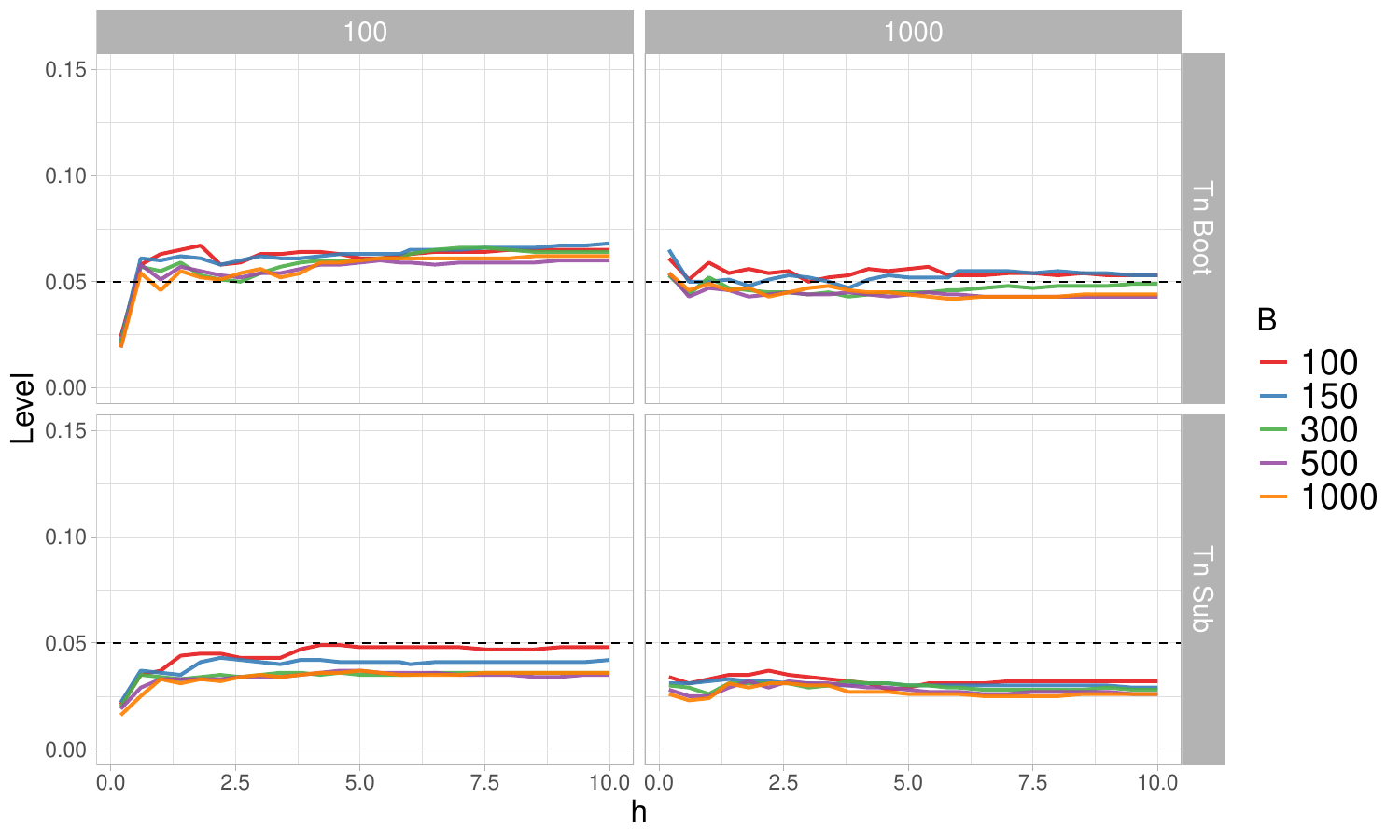}
    \caption{Level of KBQD tests for increasing $h$ with respect to the bootstrap and subsampling algorithm, for different values of $n$, $N = 1000$ replications, $d=2$. The two samples are generated from the $d$-dimensional standard normal distribution. The dashed line denotes the nominal level $\alpha = 0.05$.}
    \label{fig:level_B}
\end{figure}
Figure \ref{fig:boxplot-B} shows the boxplots of critical values computed with the considered algorithms for $h=2.2$ and different values of $n$ and $d$, indicated as headers. The computed critical values show less variability for larger $B$, while there are no evident differences in terms of sample size and dimension.

Finally, we investigate whether the KBQD tests are affected by the value of $B$, in terms of level and power. 
Figure \ref{fig:level_B} compares the level achieved by the KBQD tests with respect to $B$, for increasing $h$, $d=6$ and different values of $n$, indicated as header. In general, subsampling achieves a lower level than the nominal level, preferring lower values of $B$, especially in low dimensions. For bootstrap and permutation, increasing the number of replications $B$ improves the performance in terms of level, while for large $d$ the differences are less evident and the achieved level is equal to the nominal one for almost all values of $h$.
\begin{figure}[!htb]
    \centering
    \includegraphics[width=.95\textwidth]{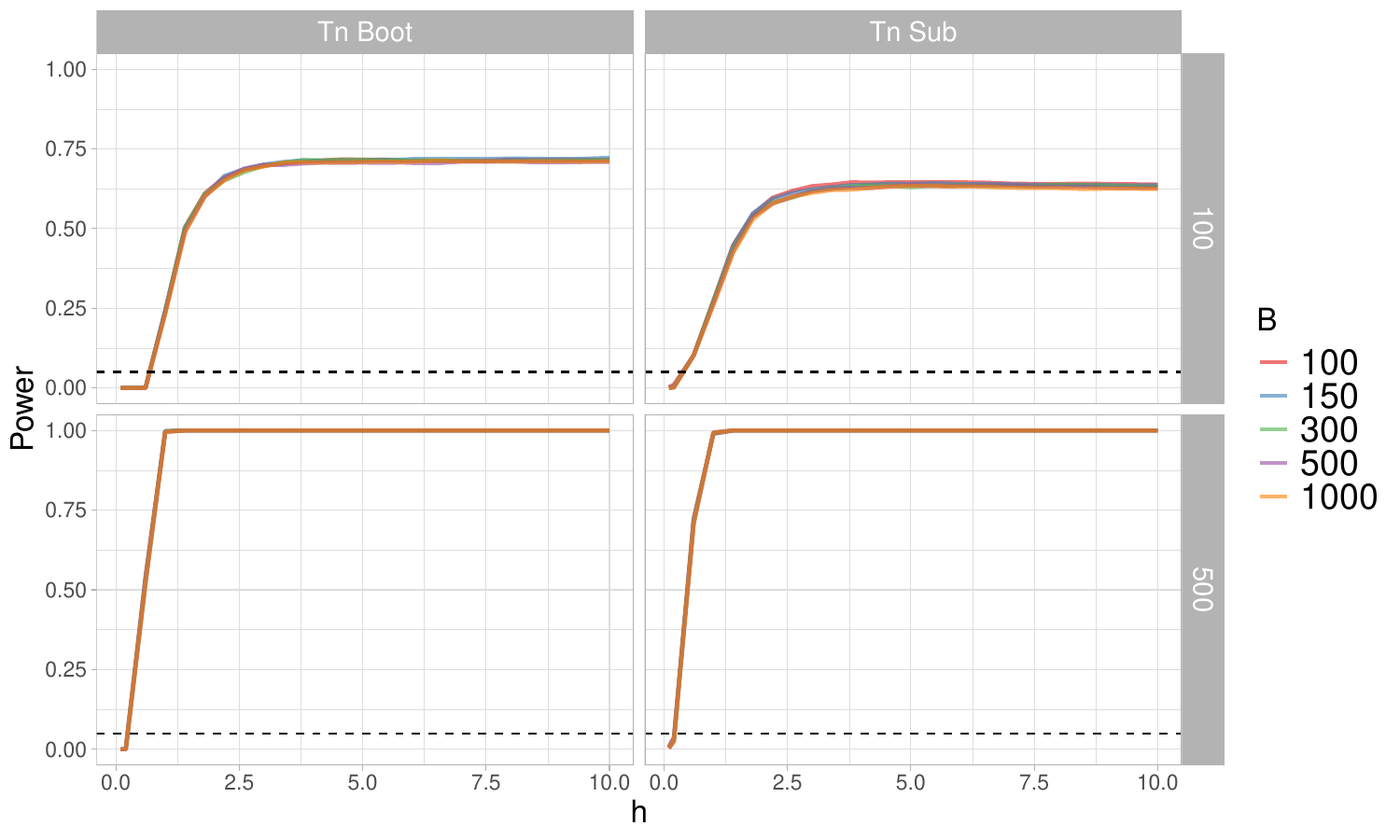}
    \caption{Power of KBQD tests for increasing $h$ with respect to the bootstrap, permutation and subsampling algorithm. One sample is generated from the $d$-dimensional standard normal distribution, while the second sample is generated from the skew-normal distribution $SN_d(\vect{0},\mat{I}_d,\vect{\lambda})$, with $\vect{\lambda} = 0.3$, for $n=100, 500$, $N = 1000$ replications and $d=6$. The dashed line denotes the nominal level $\alpha = 0.05$.}
    \label{fig:power_B}
\end{figure}
Figure \ref{fig:power_B} shows the power of the KBQD tests for the subsampling and bootstrap algorithms and the different values of $B$, when $n=100, 500$, $d=6$ and $\lambda=0.3$. For small sample size, the tests with lower number of replications achieve slightly higher power while for the other combinations of parameter values the obtained power does not change significantly. 

In summary, permutation and bootstrap sampling methods are preferred in high-dimensional spaces and small sample sizes due to their stability in performance, while subsampling is favored for large sample sizes for its lower computational time. 
Considering the evaluation of the performances of the KBQD tests with respect to the number of replications $B$ reported in this section, we recommend the usage of $B=150$ as default setting, offering a balance between statistical power, stability and computational feasibility. 

\begin{figure}[!htb]
    \centering
    \includegraphics[width=\textwidth]{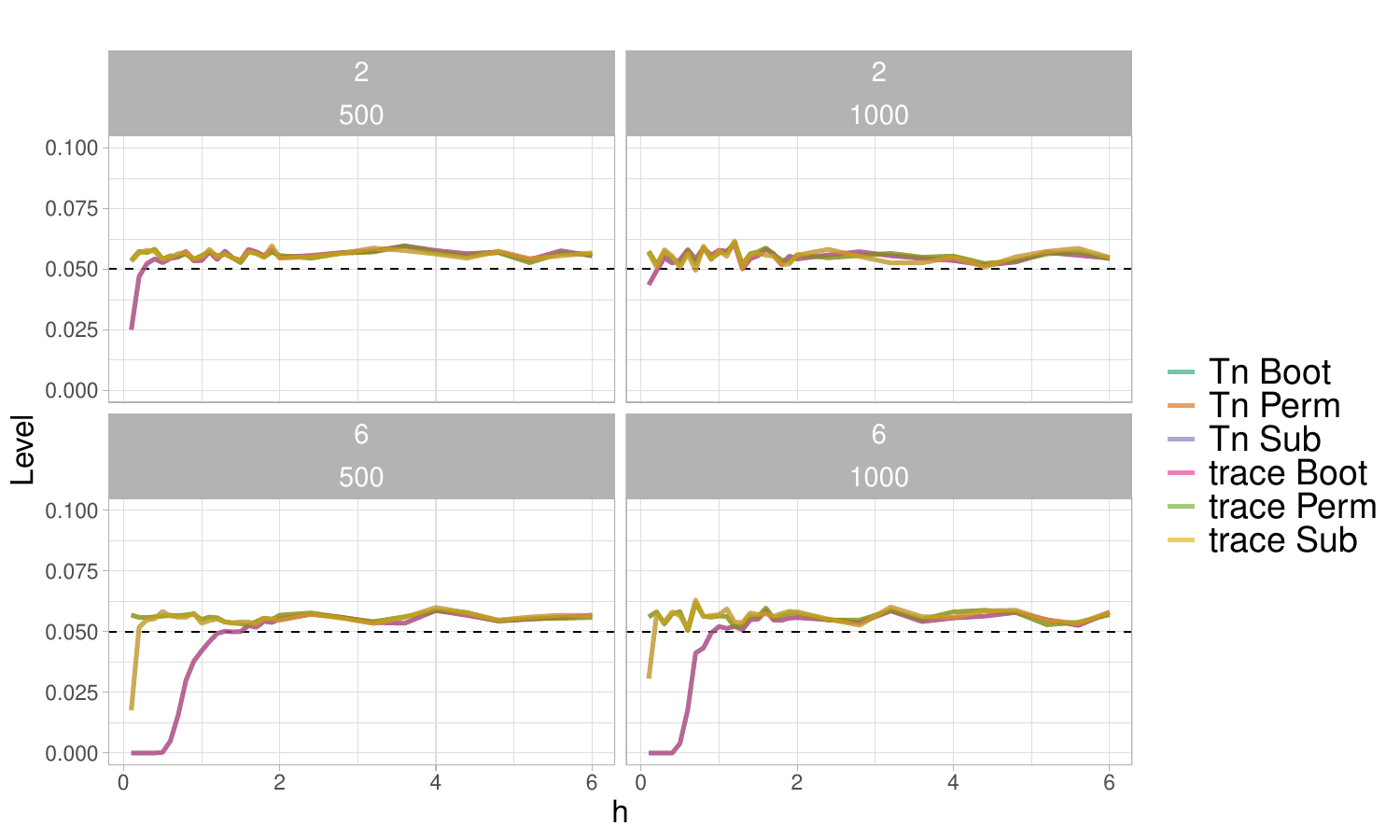}
    \caption{Level of KBQD tests for increasing $h$ with respect to the bootstrap, permutation and subsampling algorithm. The two samples are generated from the $d$-dimensional standard normal distribution, for $n=500, 1000$, $d=2,6$ and $N = 10000$ replications. The dashed line denotes the nominal level $\alpha = 0.05$.} 
    \label{fig:level_B_N}
\end{figure}

Considering the performance in terms of level in Figure \ref{fig:level_B}, we increased the number of replications to $N=10000$, for dimension $d=2,6$ and sample size $n=500, 1000$. The KBQD tests are computed for $B=150$ and $h = 0.2(0.4)6$. Figure \ref{fig:level_B_N} presents the level of the $T_n$ and $\mathrm{trace}$ statistics for each combination of dimension and sample size.  
The results indicate that the KBQD tests consistently maintain the nominal level regardless of the algorithm used to compute the critical value. 

\subsection{Selection of tuning parameter h}
\label{subsec:selection-h}

In the one-sample problem, \cite{Lindsay2014} proposed a strategy to select the tuning parameter of the kernel used to build the quadratic distances comparing the sensitivity of a class of kernels indexed by the tuning parameter. We properly adapt this procedure in order to select the appropriate tuning parameter for the k-sample test statistics. 

The distribution theory of the test statistics is rather complex, so in order to study the performance characteristics of the tests we resort to simulation. To select the tuning parameter $h$ we use a grid search algorithm.
First, we select a family of target alternatives, denoted as $\{F_\delta\}$, with $\delta \ge 0$. Here, $\delta = 0$ corresponds to the null hypothesis, and increments in $\delta$ signify increasingly pronounced deviations from $H_0$.
As a final value for $h$ we choose the minimum number that returns empirical power equal to or greater than 0.5. This value is chosen in analogy to the discussion on power considerations in \citet[Section~3]{Lindsay2014}.

From a practical point of view, it is fundamental to identify the family of alternative distributions. We consider target alternatives defined as $F_\delta = F_\delta(\hat{\vect{\mu}}, \hat{\mat{\Sigma}}, \hat{\vect{\lambda}})$, where $\hat{\vect{\mu}}, \hat{\mat{\Sigma}}$ and $\hat{\vect{\lambda}}$ indicate the location, covariance and skewness parameter estimates from the pooled sample. 
The implementation of the described procedure is reported in Algorithm \ref{alg:h-sel}. 
\begin{algorithm}[h]
\caption{Algorithm to select optimal $h$} \label{alg:h-sel}
\begin{algorithmic}[1]
\State Compute the estimates of mean $\hat{\mu}$, covariance matrix $\hat{\Sigma}$ and skewness $\hat{\lambda}$ from the pooled sample.  
\State Choose the family of alternatives $F_\delta = F_\delta(\hat{\mu},\hat{\Sigma}, \hat{\lambda})$. 
\For{$\delta$}
    \For{$h$}
	\State Generate $\vect{X}_1,\ldots,\vect{X}_{K-1}  \sim F_0$, for $\delta=0$;
 	\State Generate $\vect{X}_K \sim F_\delta$;
	\State Compute the $k$-sample test statistic between $\vect{X}_1, \vect{X}_2, \ldots, \vect{X}_K$ with kernel parameter $h$ and the critical value;
    \State Record whether the null hypothesis is rejected;
	\State Repeat lines~5-8 $N$ times. 
	\State Compute the power of the test as the proportion of rejection across the $N$ replications. If it is greater than 0.5, select $h$ as optimal value. 
    \EndFor
\EndFor
\State If an optimal value has not been selected, choose the $h$ which corresponds to maximum power. 
\end{algorithmic}
\end{algorithm}

The presented algorithm is available in the \texttt{QuadratiK} package \citep{saraceno2024goodnessoffit}. In particular, it is automatically performed when the tuning parameter $h$ is not provided by the user for the $k$-sample test, or as the standalone function \texttt{select\_h}. The implementation of the algorithm offers: $(i)$ \textit{location} alternatives, $F_\delta = SN_d(\hat{\mu} + \delta,\hat{\Sigma}, \hat{\lambda})$,with $\delta = 0.2, 0.3, 0.4$; $(ii)$ \textit{scale} alternatives, $F_\delta = SN_d(\hat{\mu} ,\hat{\Sigma}*\delta, \hat{\lambda})$, $\delta = 0.1, 0.3, 0.5$; $(iii)$ \textit{skewness} alternatives, $F_\delta = SN_d(\hat{\mu} ,\hat{\Sigma}, \hat{\lambda} + \delta)$, with $\delta = 0.2, 0.3, 0.6$.
The values of $h = 0.4, 0.8, 1.2, 1.6, 2, 2.4, 2.8$ and $N=50$ are set as default values.
This setting provides empirical guidelines for selecting $h$, however, the optimal choice depends on the specific alternative hypothesis $F_\delta$. Thus, we recommend not to rely solely on default values, but rather explore different tuning parameter choices tailored to the specific dataset. In this respect, the function \texttt{select\_h} allows the user to set the values of $\delta$ and $h$ for a more extensive grid search. 
It is suggested to follow this approach when computational resources permit.
The performance of our algorithm has been evaluated numerically, while its theoretical properties are topic of future research. In this regard, related works such as \cite{tenreiro2022} and the aggregate test by \cite{albert2022} provide examples of theoretical investigations into kernel parameter selection.

\section{Simulation Study}
\label{sec:simulation}

In this section, we conduct a simulation study to investigate the performance of the proposed multivariate $k$-sample KBQD tests. The aim of the simulation is the study of the performance characteristics of the proposed tests vis a vis the performance of the maximum mean discrepancy and other existing $k$-sample and two-sample tests. The comparisons encompass the performance in terms of level and power, as well as computational time. In order to evaluate the performance in terms of level, we generate data under the null hypothesis, which posits that all samples are drawn from the same distribution. To assess the power performance, one of the samples is generated from the family of alternative distributions. Specifically, for each scenario we run $N$ independent replications. In each replication, we generate the data, apply the tests at the nominal level $\alpha$, and record whether $H_0$ is rejected. The empirical level (under $H_0$) and power (under the specified alternative) are then estimated as the proportion of rejections across the $N$ replications.
The test statistics are based on kernels, and an important step in the computation is the selection of the kernel tuning parameter.  
Our aim is to examine how performance varies with the bandwidth; we then evaluate level and power over a predefined grid of $h$ values. Specifically, we consider $h=0.2(0.4)10$. For the comparisons with the competitor tests, the smallest value of $h$ which yields a power greater than or equal to 0.5 is selected as optimal, following the same strategy described in Section \ref{subsec:selection-h}.
All simulations were performed in R~4.2.0 on the high-performance computing facility located at the Center for Computational Research (CCR), University at Buffalo (\url{https://ubir.buffalo.edu/xmlui/handle/10477/79221}), and on a local laptop (8~cores, base~3.2\,GHz, 16\,GB RAM; Windows 64-bit).
The implementation of the proposed kernel-based tests can be found in the \texttt{QuadratiK} package \citep{saraceno2024goodnessoffit}, available both in \textsf{R} and \textsf{Python}. 

\subsection{Two-sample testing: Asymmetric Distributions}

The proposed two-sample KBQD tests are compared with the existing multivariate two-sample tests. Our preference is selecting the two-sample tests with readily accessible implementations. These include the Friedman-Rafsky Kolmogorov-Smirnov test (FR-KS), the modified Friedman-Rafsky Kolmogorov-Smirnov test (mod-FR-KS), the Friedman-Rafsky Wald-Wolfowitz test (FR-WW), Rosenbaum's Cross Match test (Crossmatch), the energy test and the Maximum Mean Discrepancy (MMD).  Table \ref{tab:test-pack} reports the mentioned two-sample tests and the corresponding \texttt{R} packages. 
\begin{table}[ht]
    \centering
    \caption{Considered two-sample tests and corresponding R packages.}
    \label{tab:test-pack}
    \begin{tabular}{lc}
    \hline
    Test &  R package \\
    \hline
	Crossmatch & \texttt{crossmatch}\\
	Friedman-Rafsky Kolmogorov-Smirnov (FR-KS) & \texttt{GSAR}\\
    modified FR-KS & \texttt{GSAR}\\
	Friedman-Rafsky Wald-Wolfowitz (FR-WW) & \texttt{GSAR}\\
        Energy statistics & \texttt{energy}\\
	Minimum Mean Discrepancy (MMD)     &  \texttt{kernalb}\\
    \hline 
    \end{tabular}
\end{table}

The simulation study presented in section \ref{sec:computation} shows that the proposed test statistics, $\mathrm{trace}$ and $T_n$, have almost identical performance. Therefore, we will present the results for the $T_n$ statistic, with the MMD and the energy statistics used for comparison. 

In this study, we focus on the family of asymmetric alternatives. As first case, we consider samples from skew-normal distributions, denoted as $SN_d(\vect{\mu},\mat{\Sigma},\lambda)$ with location vector $\vect{\mu}$, covariance matrix $\mat{\Sigma}$ and shape parameter $\lambda$, as introduced by \cite{Azzalini1996}. 
The multivariate skew-normal distribution represents a mathematically tractable extension of the multivariate normal density, with the additional parameter $\lambda$ to regulate the skewness. When $\lambda = 0$, this corresponds to the multivariate normal. \cite{Hill1982} offer a discussion and numerical evidence supporting the presence of skewness in real data. 
Overall, the KBQD tests exhibit a level close to the nominal level in a wide range of $h$. Power increases with distributional skewness and for larger $h$.  KBQD tests with subsampling yield slightly reduced power for alternatives close to null but align with permutation/bootstrap as the skewness $\lambda$ increases, while KBQD tests with permutation and bootstrap show the strongest performance, generally outperforming the MMD and energy tests. The Crossmatch, FR-WW, and FR-KS tests perform poorly against asymmetric alternatives, with FR-KS being competitive only in the univariate case.
Complete results, including all the considered tests, are reported in section S7 of the Supplementary Material.

In this subsection, we further investigate the performance of the introduced KBQD tests against asymmetric alternatives by considering additional skewed distributions. In particular, we generate samples from the univariate Gumbel distribution $\mathrm{Gumbel}(\mu, \sigma)$, using the function \texttt{rgumbel} in the \texttt{R} package \texttt{evd}, and the univariate log-normal distribution using the function \texttt{rlnorm} in the \texttt{R} package \texttt{stats},  both depending on two parameters $\mu$ and $\sigma$ which denote the location and scale parameters. We generate univariate observations according to the following scenarios. 
\begin{itemize}
    \item \textbf{Log-normal}:  $\vect{x}_1,\ldots,\vect{x}_{n_1} \sim \mathrm{log}N(0,0.8)$ and $\vect{y}_1,\ldots,\vect{y}_{n_2} \sim \mathrm{log}N(0,\sigma)$ with $\sigma=0.6, 0.65, 0.7, 0.75, 0.8$.
    \item \textbf{Gumbel-scale}: $\vect{x}_1,\ldots,\vect{x}_{n_1} \sim \mathrm{Gumbel}(0,1)$ and $\vect{y}_1,\ldots,\vect{y}_{n_2} \sim \mathrm{Gumbel}(0,\sigma)$ with $\sigma=0.8, 0.9, 1, 1.1, 1.2$.
    \item \textbf{Gumbel-location}: $\vect{x}_1,\ldots,\vect{x}_{n_1} \sim \mathrm{Gumbel}(0,1)$ and $\vect{y}_1,\ldots,\vect{y}_{n_2} \sim \mathrm{Gumbel}(\mu,1)$ with $\mu=0, 0.1, 0.2, 0.3$.
\end{itemize}
The sample size is $n_1=n_2=50,100,500$, $B=150$, and the number of replications $N=1000$. For these cases, we compare the KBQD tests only with MMD and energy tests, considering the poor performance of the other tests in the previous simulations with samples generated from the skew-normal distribution. 
The considered families of alternatives are displayed in Figure S20 and Figure S21, respectively, Section S7.5 of Supplementary Material.

\begin{figure}[htb]
    \centering
    \includegraphics[width=0.8\textwidth]{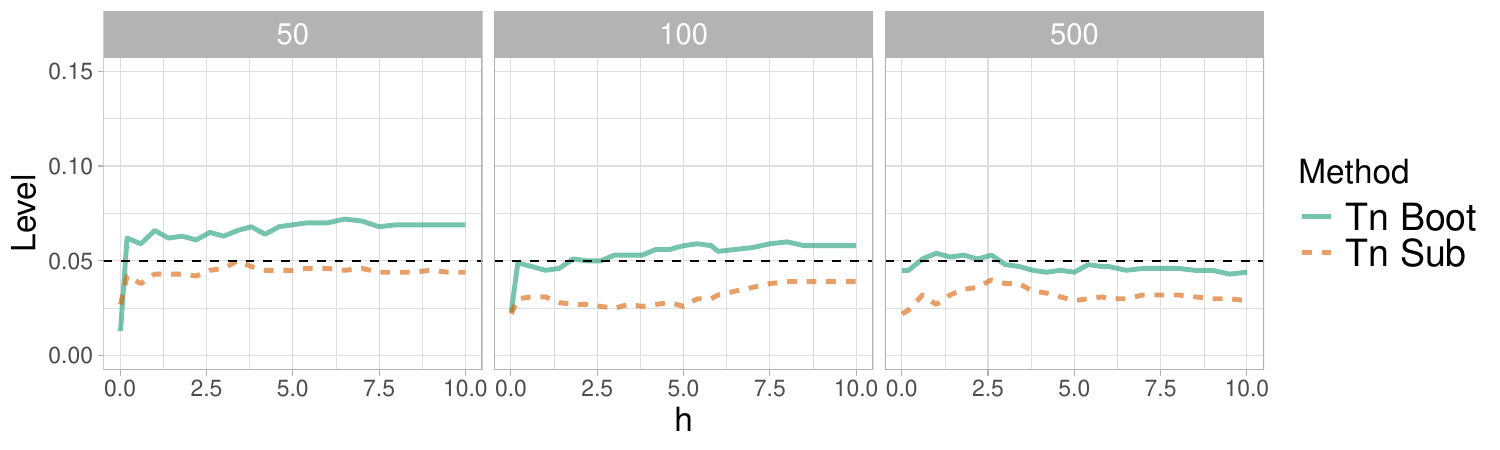}
    \includegraphics[width=0.8\textwidth]{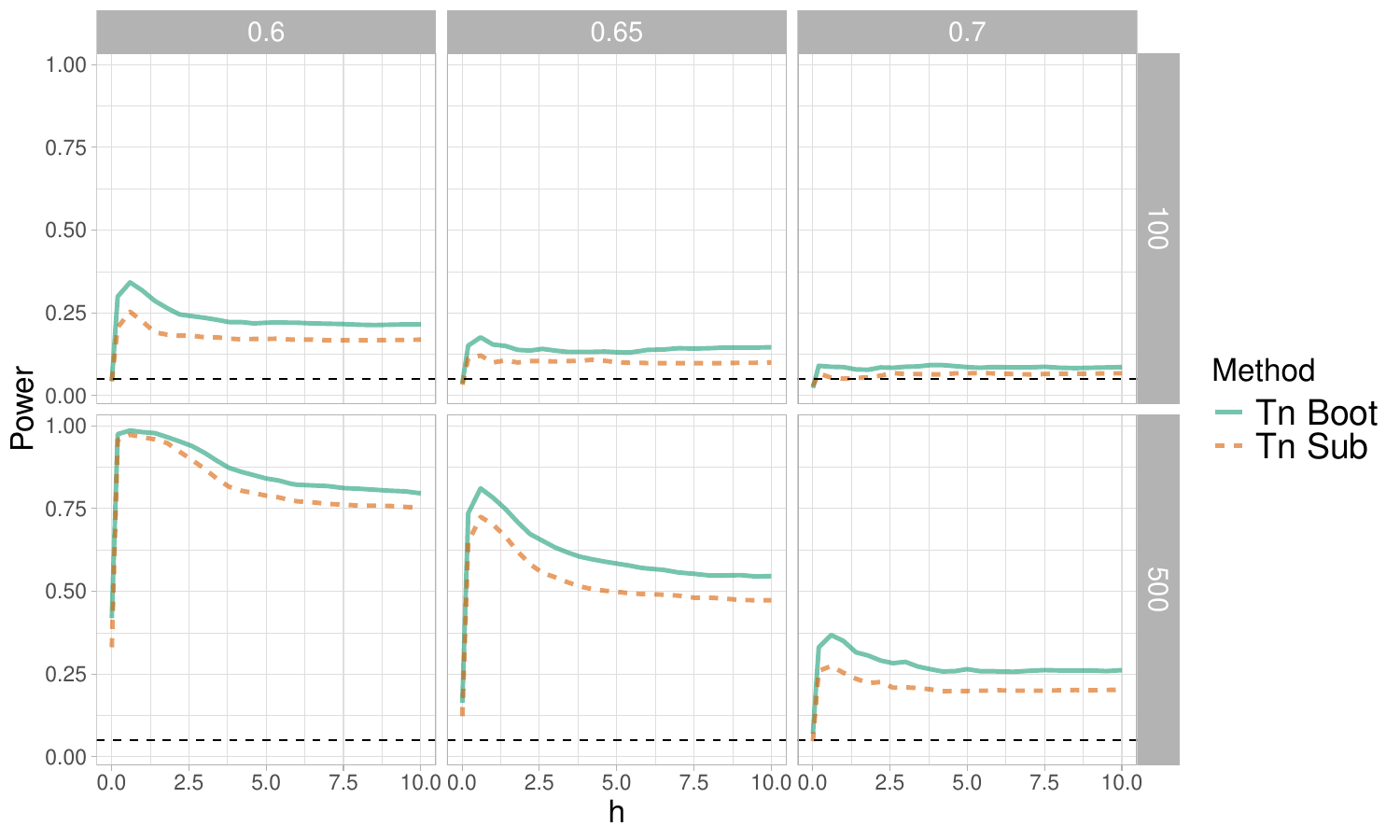}
    \caption{One sample is generated from the lognormal distribution $\mathrm{log}N(0, 0.8)$, while the second sample from $\mathrm{log}N(0, \sigma)$. (top) Level of the kernel-based tests, computed using the bootstrap and subsampling algorithm, with respect to the parameter $h$ for $\sigma = 0.8$. (bottom) Power of the kernel-based tests, computed using the bootstrap and subsampling algorithm, with respect to the parameter $h$ for for different values of $\sigma$ and samples size, indicated as headers.}
    \label{fig:power-logn-scale}
\end{figure}
\begin{figure}[!htb]
    \centering
    \includegraphics[width=\textwidth]{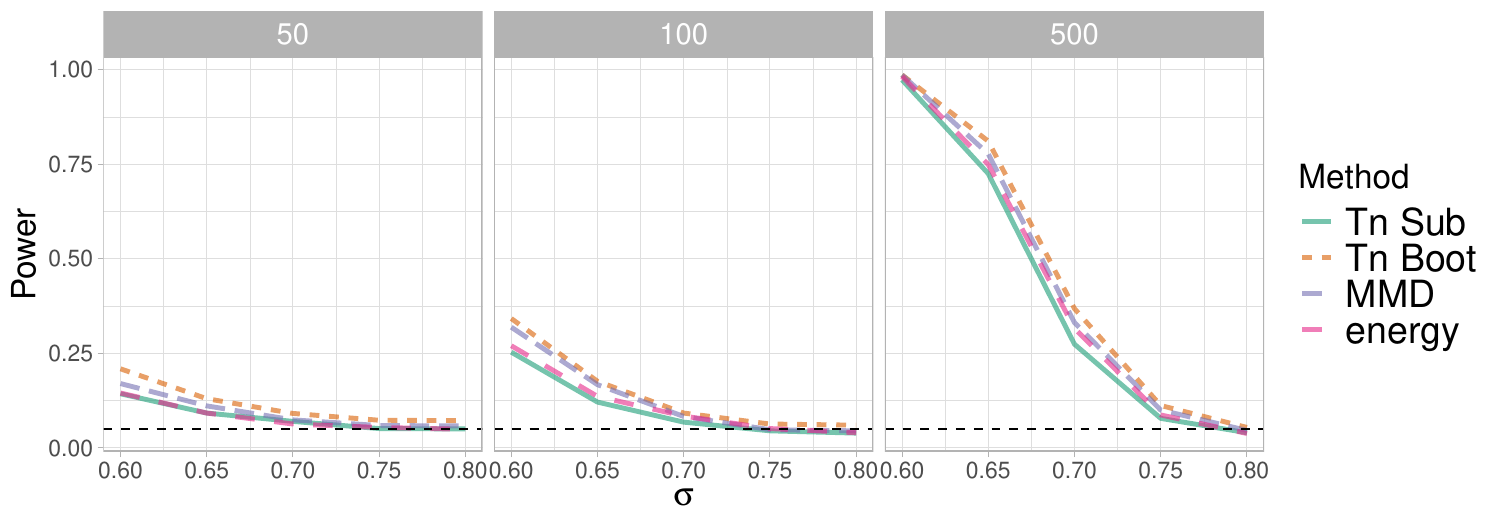}
    
    \caption{One sample is generated from the lognormal distribution $\mathrm{log}N(0, 0.8)$, while the second sample from $\mathrm{log}N(0, \sigma)$. Power of the considered tests with respect to different values of $\sigma$, in the $x$-axis, and samples size, indicated as header. The value of $h$ is selected via the grid-search algorithm described in Section \ref{subsec:selection-h}.}
    \label{fig:power-logn-sigma-all}
\end{figure}
\begin{figure}[htb]
    \centering
    \includegraphics[width=.95\textwidth]{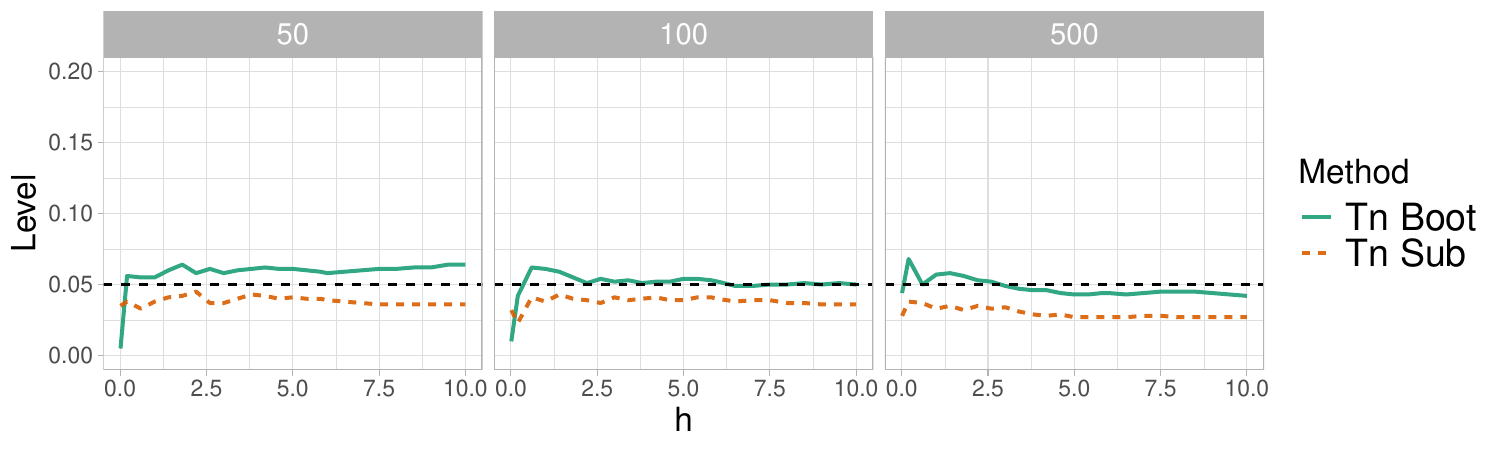}
    \includegraphics[width=.95\textwidth]{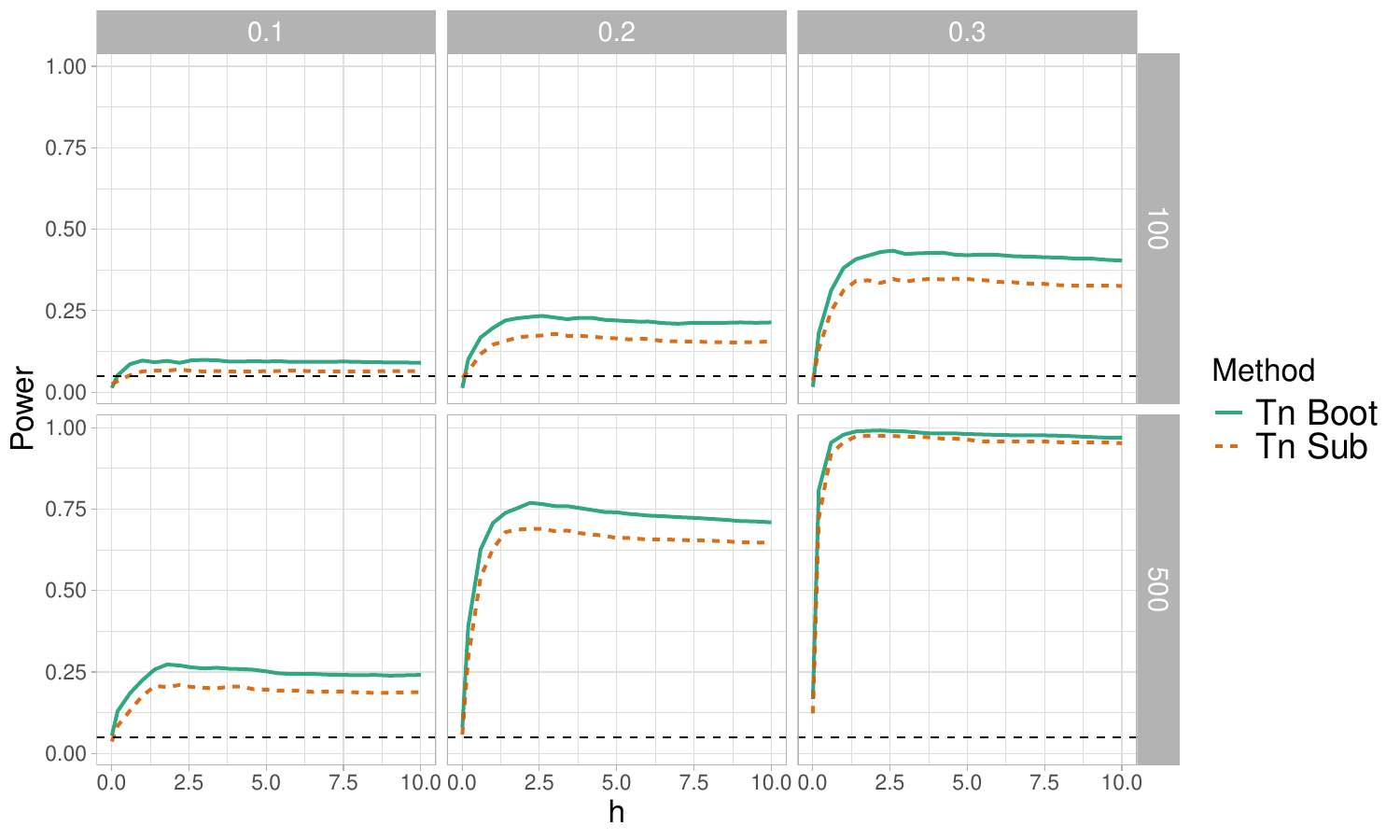}
    \caption{One sample is generated from the $\mathrm{Gumbel}(0, 1)$ distribution, while the second sample from $\mathrm{Gumbel}(\mu, 1)$. (top) Level of the kernel-based tests, computed using the bootstrap and subsampling algorithm, with respect to the parameter $h$ for 
    $\mu = 0$. (bottom) Power of the kernel-based tests, computed using the bootstrap, permutation and subsampling algorithm, with respect to the parameter $h$ for 
    different values of $\mu$ and samples size, indicated as headers.}
    \label{fig:power-gumbel-mean}
\end{figure}
\begin{figure}[!htb]
    \centering
    \includegraphics[width=\textwidth]{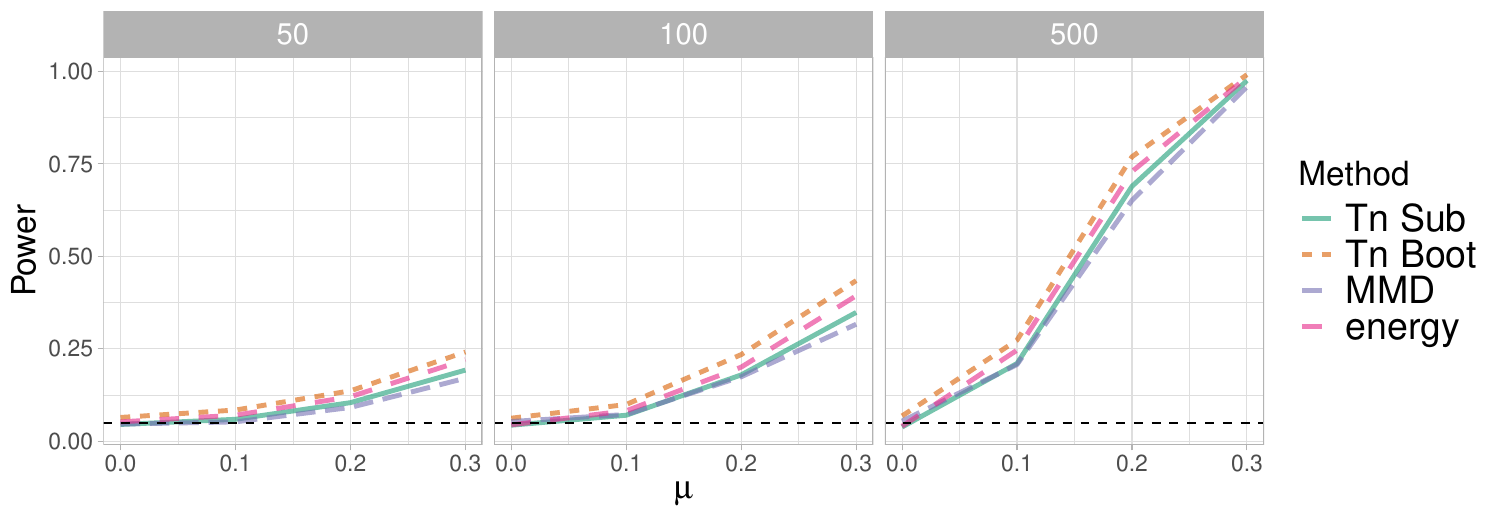}
    
    \caption{One sample is generated from the $\mathrm{Gumbel}(0, 1)$ distribution, while the second sample from $\mathrm{Gumbel}(\mu, 1)$. Power of the considered tests for different values of $\mu$, in the $x$-axis, and samples size, indicated as header. The value of $h$ is selected via the grid-search algorithm described in Section \ref{subsec:selection-h}.
    }
    \label{fig:power-gumbel-mean-all}
\end{figure}
\begin{figure}[htb]
    \centering
    \includegraphics[width=0.95\textwidth]{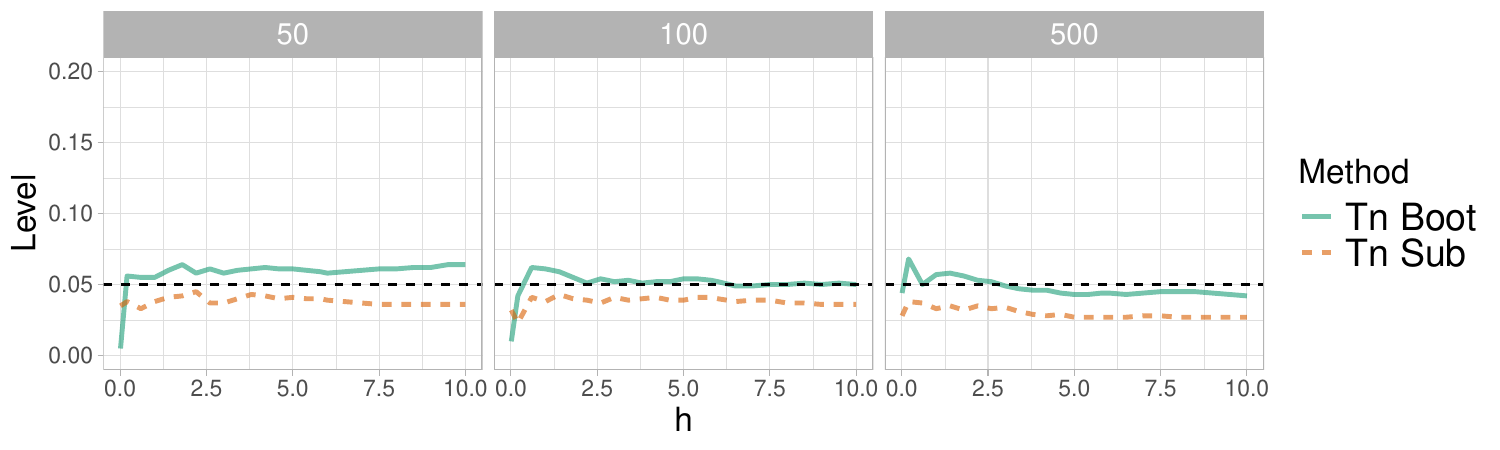}
    \includegraphics[width=0.95\textwidth]{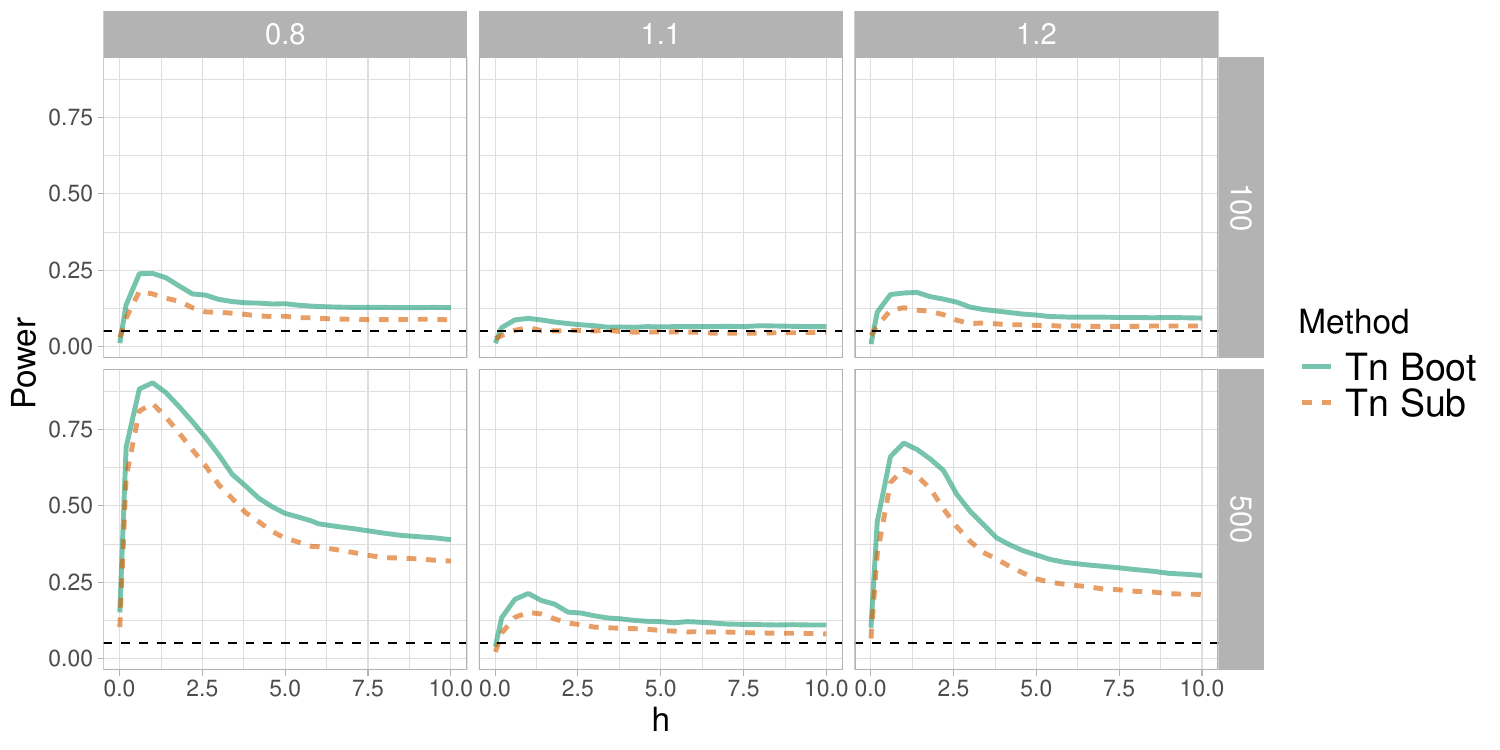}
    \caption{One sample is generated from the $\mathrm{Gumbel}(0, 1)$ distribution, while the second sample from $\mathrm{Gumbel}(0, \sigma)$. (top) Level of the kernel-based tests, computed using the bootstrap and subsampling algorithm, with respect to the parameter $h$ for $\sigma = 0$ and samples size indicated as header. (bottom) Power of the kernel-based tests, computed using the bootstrap, permutation and subsampling algorithm, with respect to the parameter $h$ for different values of $\sigma$ and samples size, indicated as headers.}
    \label{fig:power-gumbel-scale}
\end{figure}

Figure~\ref{fig:power-logn-scale}, Figure~\ref{fig:power-gumbel-mean} and Figure~\ref{fig:power-gumbel-scale} show the performance of the proposed kernel-based tests, using the bootstrap and subsampling algorithms, with respect to the tuning parameter $h$ for the described scenarios. When the alternative distribution is close to the null hypothesis, the choice of $h$ does not affect the power of the resulting test. When the departure from the null hypothesis is more evident, a range of values is suggested for obtaining powerful tests. In particular, for the Gumbel-scale scenario, the simulation results suggest that all values greater than a certain threshold can be chosen for optimal power (Figure~\ref{fig:power-gumbel-mean}), while for the other two scenarios the optimal choices reduce to an interval (Figure~\ref{fig:power-logn-scale} and Figure~\ref{fig:power-gumbel-scale}), showing that for increasing $h$ the power decreases.
\begin{figure}[!htb]
    \centering
    \includegraphics[width=\textwidth]{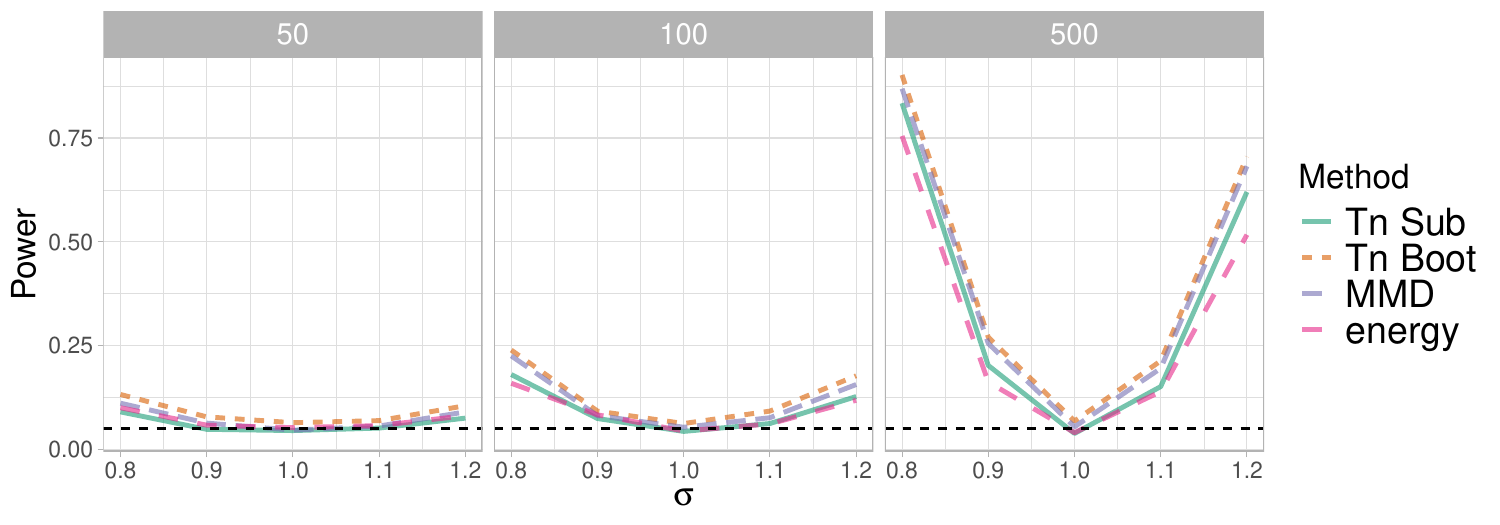}
    \caption{One sample is generated from the $\mathrm{Gumbel}(0, 1)$ distribution, while the second sample from $\mathrm{Gumbel}(0, \sigma)$. Power of the considered tests for different values of $\sigma$, in the $x$-axis, and samples size, indicated as header. The value of $h$ is selected via the grid-search algorithm described in Section \ref{subsec:selection-h}.}
    \label{fig:power-gumbel-sigma-all}
\end{figure}
Figure~\ref{fig:power-logn-sigma-all} shows the performance in terms of power of the KBQD tests in comparison to the MMD and energy tests, when the $\log$-normal distribution is considered. Figure~\ref{fig:power-gumbel-mean-all} and Figure~\ref{fig:power-gumbel-sigma-all} display the achieved power of the considered two-sample tests when the family of alternatives follows the Gumbel distribution, according to scenarios Gumbel-scale and Gumbel-location, respectively.   
The KBQD tests with the bootstrap and permutation algorithms show an improved power with respect to MMD and energy tests, especially for low sample size. The two kernel-based tests achieve the same performance for increasing sample size and departure from the null hypothesis. 

\subsection{\textit{k}-Sample Problem}

We investigate the performance of the $k$-sample KBQD tests in the case of $k$ samples. The existing multivariate $k$-sample tests are considered for analyzing the differences with the proposed tests. 
The literature includes few multivariate $k$-sample tests with a readily available implementation. In this simulation study, we include the $k$-sample energy and the independence $k$-sample tests, available in the \texttt{Python} package \texttt{hyppo}. 
The \texttt{KSample} function in the \texttt{hyppo} package is performed using as independence tests the distance correlation \texttt{DCorr}, which is equivalent to the $k$-sample Energy test, and the Hilbert-Schmidt independence criterion \texttt{Hsic}, which corresponds to the MMD. 

\subsubsection*{The Bivariate Normal Case}

We consider the hypothesis testing problem $H_0: F_1 = F_2 = F_3$ with $F_i \sim N_2(\vect{\mu},\mat{I}_2)$ follows a bivariate normal distribution with mean vector $\vect{\mu}$. We distinguish three cases.
\begin{itemize}
    \item \textbf{None}: The three samples are generated from the same normal distribution with $\vect{\mu}=(0,0)$.
    \item \textbf{One}: Two samples are generated from the standard normal distribution, and $F_3 \sim N_2(\vect{\mu},\mat{I}_2)$ with $\vect{\mu}=(0,\epsilon)$.
    \item \textbf{All}: The three samples are generated from different normal distributions with mean vectors $\vect{\mu}_1=(0,\sqrt{3}/3 \times \epsilon)$, $\vect{\mu}_2=(-\epsilon/2,-\sqrt{3}/6 \times \epsilon)$ and $\vect{\mu}_3=(\epsilon/2,-\sqrt{3}/6 \times \epsilon)$.
\end{itemize}
For all these cases, we consider sample size $n_1=n_2=n_3=100,200,500$, $\epsilon = 0,0.1,\ldots,1$, $B=150$ and $N=1000$ replications.
\begin{figure}[htb]
    \centering
    \includegraphics[width=.6\textwidth]{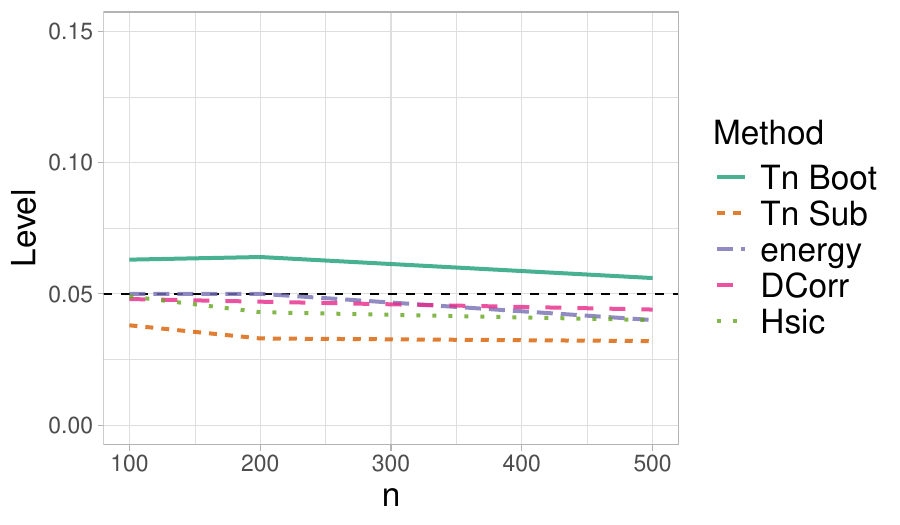}
    \caption{Three samples are generated from the $d$-dimensional normal distribution. Level of the $k$-sample KBQD test $T_n$, with the critical value computed using bootstrap and subsampling, compared with the available $k$-sample tests, as function of sample size $n$ and $d=2$. The dashed line corresponds to the nominal level 0.05.}
    \label{fig:level-gauss-k}
\end{figure}
\begin{figure}[!ht]
    \centering
    \includegraphics[width=\textwidth]{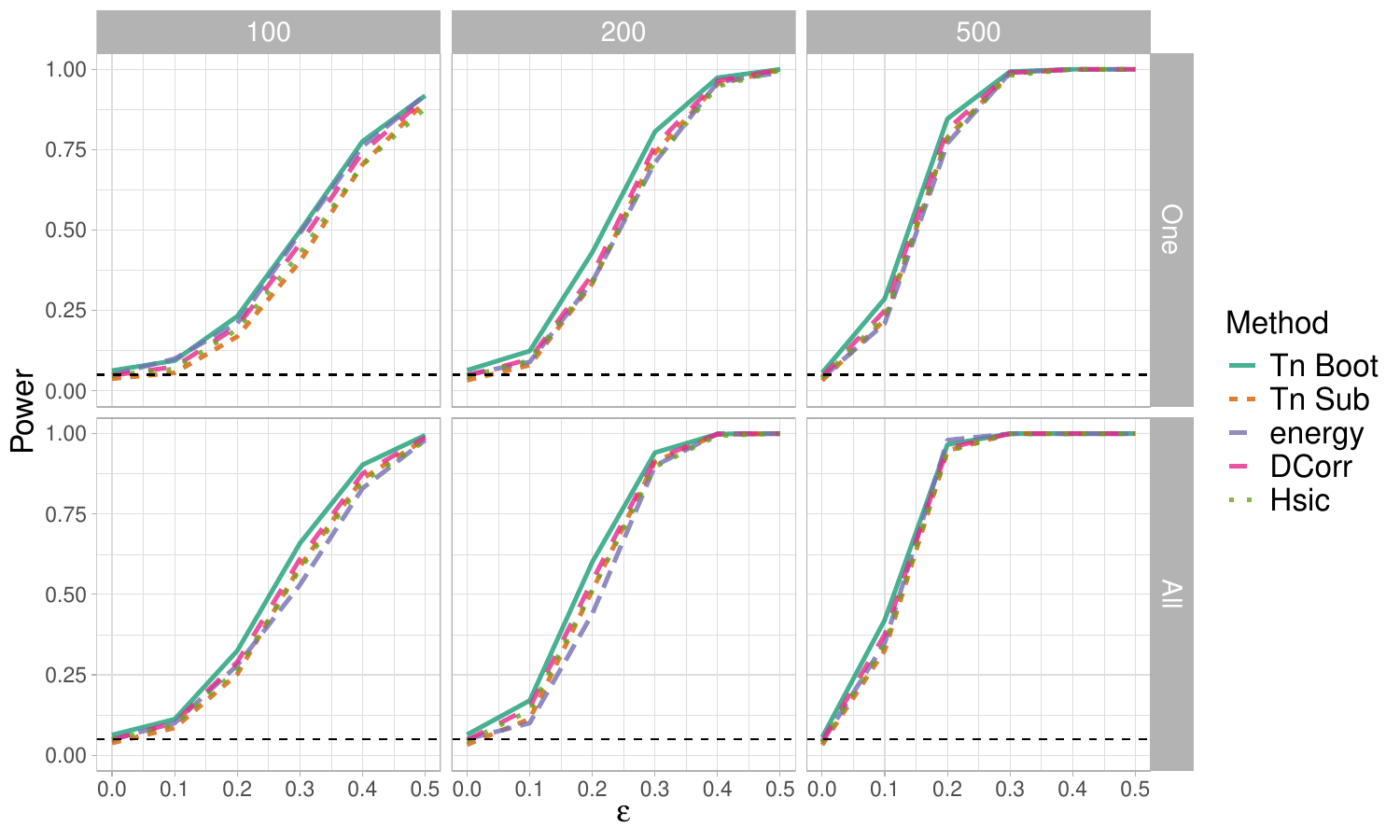}
    \caption{Normal distribution: Power of the KBQD test $T_n$, with the critical value computed using bootstrap and subsampling, as function of $\epsilon$ for sample size $n=100,200,500$ indicated as column headers. The simulation scenario is indicated as row header. We display here $\epsilon \in [0,0.5]$.}
    \label{fig:power-gauss-k}
\end{figure}
Figure \ref{fig:level-gauss-k} shows the level of the KBQD test compared to the energy test and the independence tests, in the \textbf{None} case, in which the null hypothesis is true. The power achieved by the considered tests is displayed in Figure \ref{fig:power-gauss-k} for the other two cases and different sample sizes. The proposed KBQD tests, with bootstrap and permutation algorithms, show slightly higher power than the other tests especially for lower sample size and small departures from the null hypothesis, maintaining a level close to the nominal one. The KBQD test with the subsampling algorithm shows similar performance to the MMD.  

\subsubsection*{Additional alternative distributions}

We focus now on families of alternative distributions different from the Gaussian distribution. More specifically, we generate $k-1$ samples according to the null hypothesis, and the remaining sample from the family of alternatives indexed by $\epsilon$. The following alternatives are taken into consideration.
\begin{itemize}
    \item \textbf{Cauchy distribution}: One sample is generated from the $d$-variate Cauchy distribution, that is $F_1 \sim \mathrm{Cauchy}(\vect{\varepsilon},\mat{I}_d)$, while $F_i \sim \mathrm{Cauchy}(\vect{0},\mat{I}_d)$, for $i \not=1$.
    \item \textbf{$t$-distribution}: One sample is generated from the $d$-variate $t$ distribution with 4 degrees of freedom, that is $F_1 \sim t_4(\vect{\varepsilon},\mat{I}_d)$, while $F_i \sim t_4(\vect{0},\mat{I}_d)$, for $i \not=1$.  
\end{itemize}
We consider $k=3,5$ number of samples, $B=150$, $N=1000$, $n=50,200,500$ per sample with $d=2,6,10$. To investigate the level of the considered $k$-sample tests, we include larger sample sizes $n=1000, 3000$. To assess the performance of the tests in terms of power, for each family of distributions, we consider two types of alternatives:
\begin{itemize}
    \item[]Type 1: $\vect{\varepsilon}$ is a $d$-dimensional constant vector, that is $\vect{\varepsilon} = (\epsilon,\ldots,\epsilon)$.
    \item[]Type 2: $\vect{\varepsilon}$ has the first half elements equal to $\epsilon$ and the remaining equal to zero.
\end{itemize}
We report here the simulation results for alternatives following the Cauchy distribution, while the results for the $t$-distribution can be found in Section S8 of the Supplementary Material.

\begin{figure}[!ht]
    \centering
    \includegraphics[width=\textwidth]{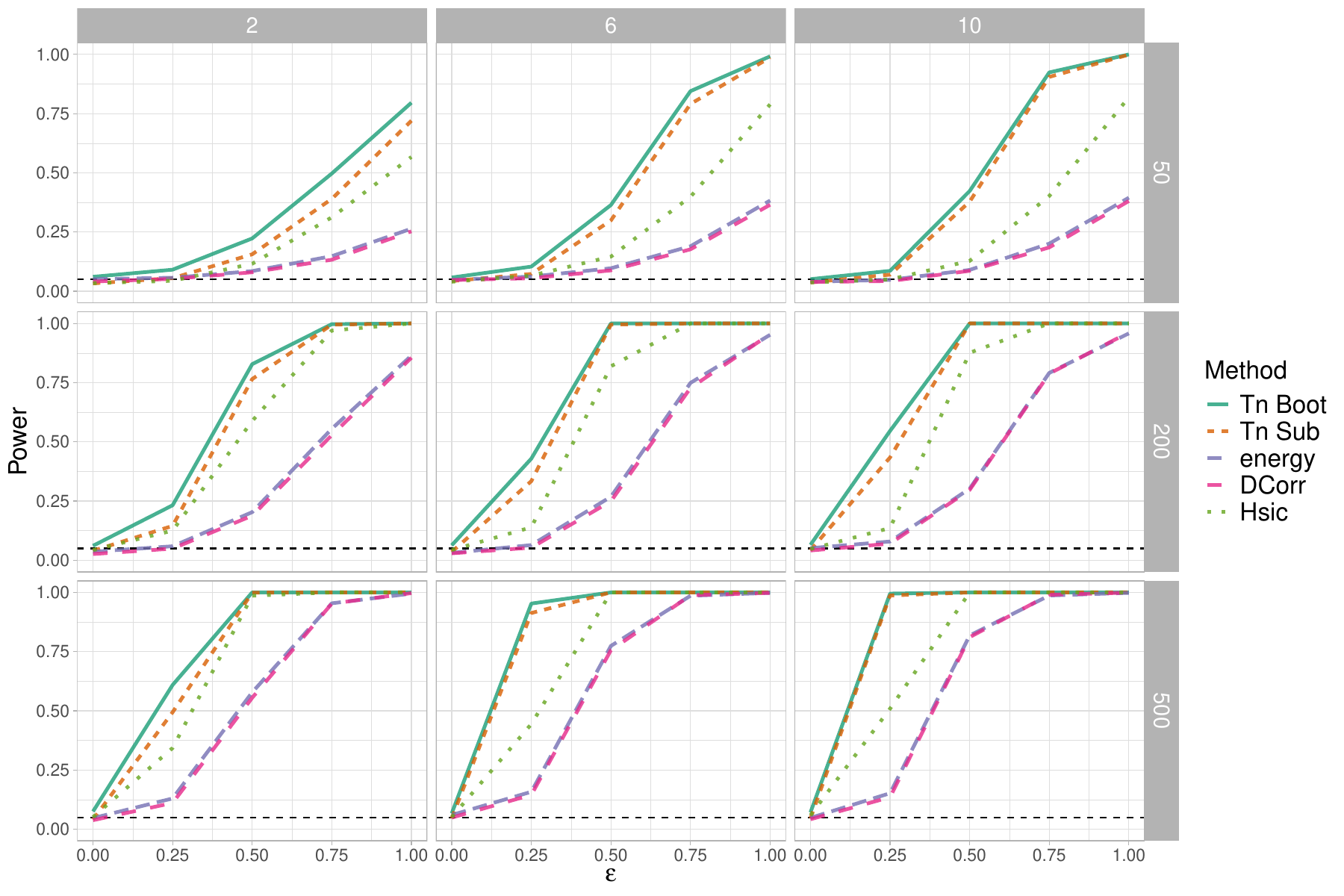}
    \caption{Cauchy distribution: Power of KBQD test $T_n$, with the critical value computed using bootstrap and subsampling, compared with the available $k$-sample tests, as function of $\varepsilon$ following \textbf{Type 2} alternatives, with $\epsilon$ from 0 to 1, for sample size per group $n=50,200,500$ and dimension $d=2,6,10$, indicated as headers. The number of samples is $k=5$. The dashed black line indicates the nominal level 0.05.}
    \label{fig:power-chaucy-k}
\end{figure}

Figure \ref{fig:level-chaucy-k} depicts the level obtained by the KBQD tests, the energy test and the independence tests when testing the equality of samples from Cauchy distributions as function of the sample size $n$, for different values of dimension $d$ and number of samples $k$. Figure \ref{fig:power-chaucy-k} compares the performance of the $k$-sample tests in terms of power as function of $\varepsilon$, for \textbf{Type 2} alternatives and $k=5$ samples, considering different values of dimension $d$ and sample size (per sample) $n$.

\begin{figure}[!t]
    \centering
    \includegraphics[width=.95\textwidth]{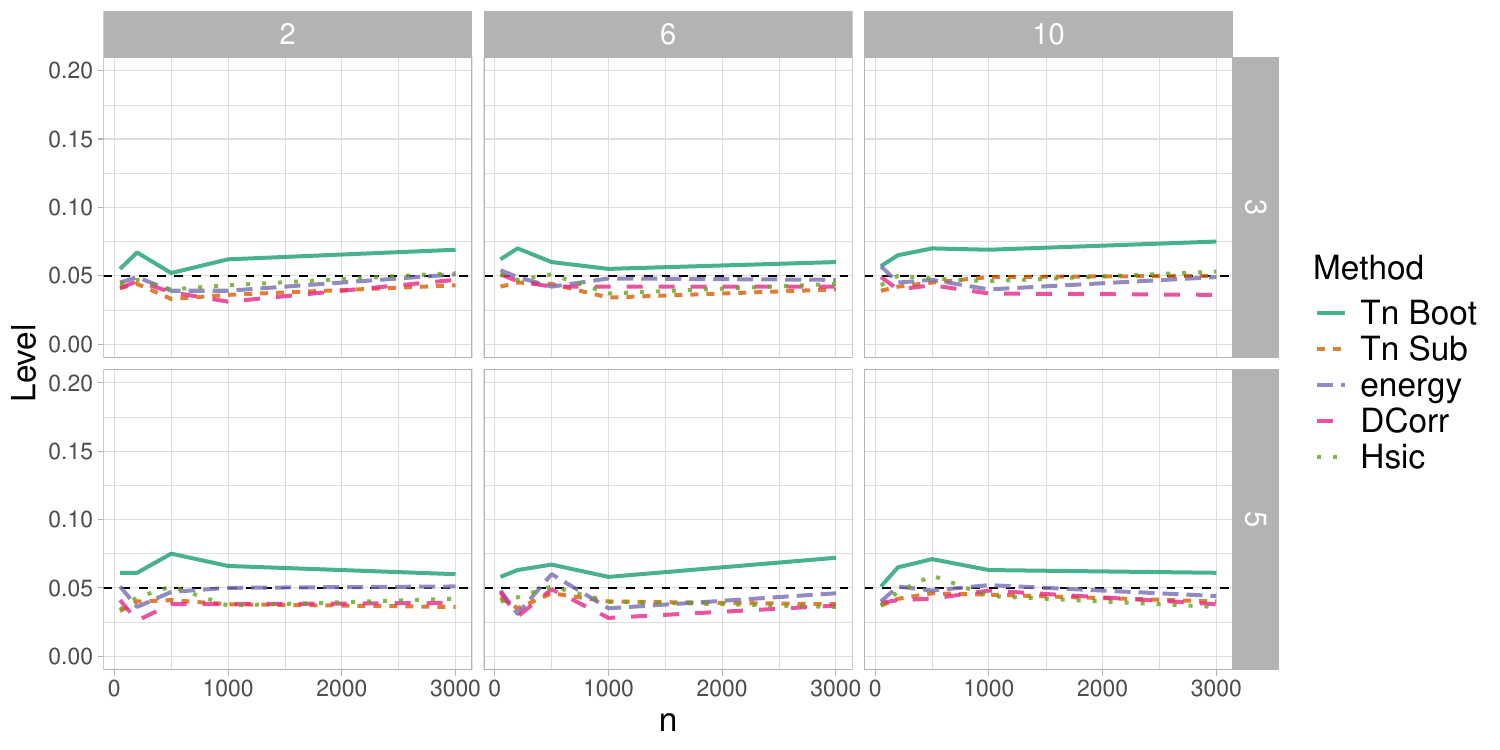}
    \caption{Cauchy distribution: Level of KBQD test $T_n$, with the critical value computed using bootstrap and subsampling, compared with the available $k$-sample tests, as function of sample size $n$, for dimension $d=2,6,10$ and number of samples $k=3,5$ indicated as headers. The dashed black line indicates the nominal level 0.05.}
    \label{fig:level-chaucy-k}
\end{figure}
The KBQD tests with the bootstrap and permutation algorithms show slightly higher level than the nominal one. When one considers the Monte Carlo error, the level obtained is approximately $0.065$, consistent with the reported level in \cite{Lindsay2014}. An increase in the number of Monte Carlo replications brings the level closer to the nominal value. On the other hand, the level of KBQD test using the subsampling algorithm is closer to the nominal level, with deviations diminishing for increasing sample size. Figure \ref{fig:power-chaucy-k} shows that the proposed KBQD tests outperform the available $k$-sample tests, especially for higher dimensions and small sample sizes. The Independence test \textit{Hsic} is competitive only for dimension $d=2$.

\subsection{Real data application}
\label{subsec:real-data}

\begin{figure}[!htb]
    \centering
    \includegraphics[width=\textwidth]{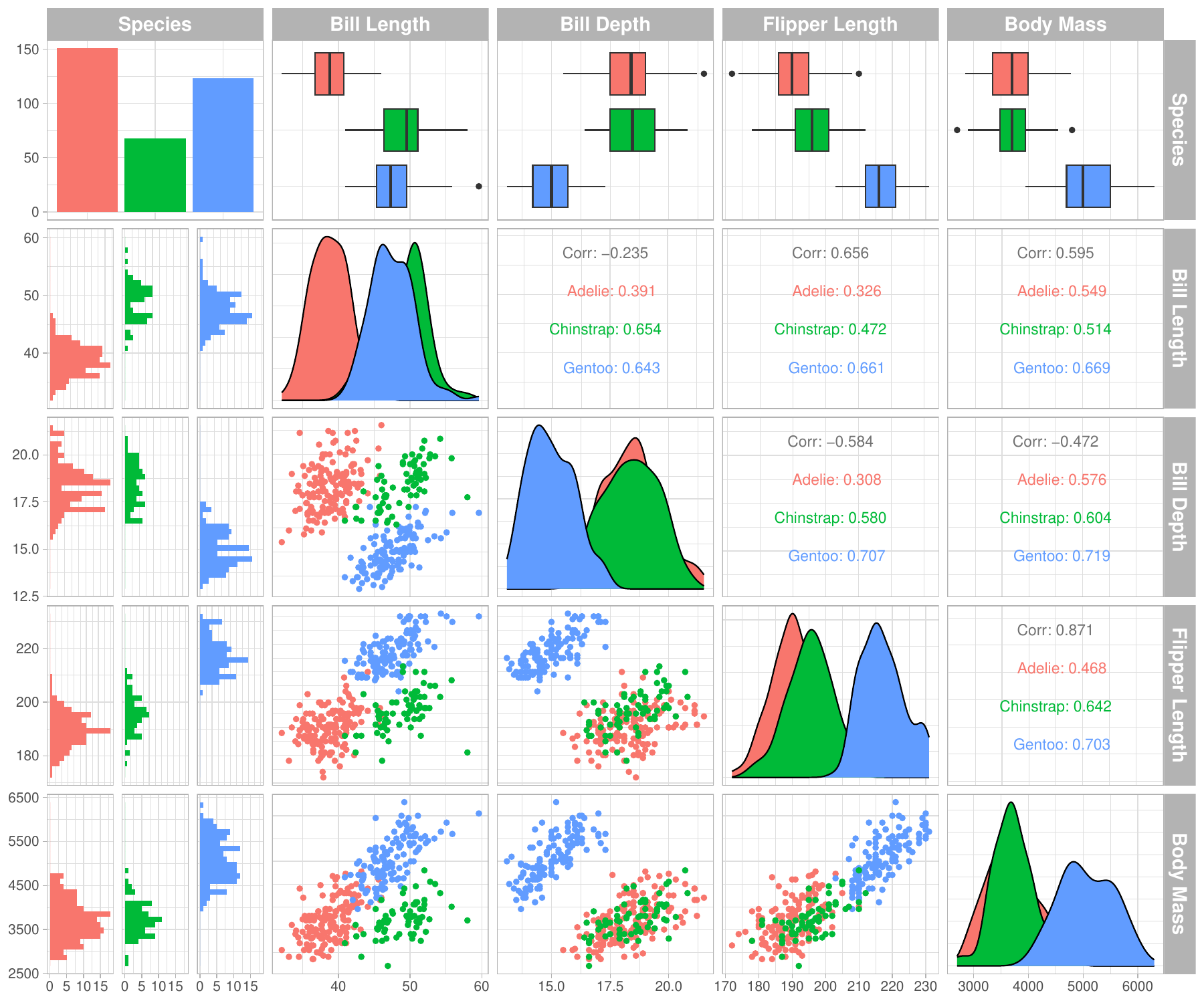}
    \caption{Pairs plots of the variables in the penguin data set. It displays the histogram and boxplots of each variable on the first column and the first row, respectively. On the main diagonal, the estimated densities are showed, while the bivariate scatter plot and pairwise correlations are reported on the off diagonal. In all the plot, the species is indicated with different colors.}
    \label{fig:peng-pairs}
\end{figure}
Here, we consider the Palmer Archipelago (Antartica) penguin dataset, originally collected by \cite{Gorman2014} and made available in the R package \texttt{palmerpenguins} \citep{penguin-data}. 
This data set contains complete measurements of 4 continuous variables for 342 penguins, which belong to 3 different species; the measurements were collected from 3 islands in the Palmer Archipelago, Antartica.  

For each variable, Figure~\ref{fig:peng-pairs} displays the estimated density, boxplot, histogram and bi-variate scatter plots colored by species.
The boxplots and histograms show that penguins from the three species present differences with respect to the considered characteristics. In particular, the estimated densities show that the ``Adelie" and ``Chinstrap" species appear to follow similar distributions in all the variables except for ``Bill Length". In fact, from the bivariate scatter plots, the species are well separated when ``Bill Length" is considered together with another variable.  

At first, we perform the kernel-based quadratic distance $k$-sample tests and several other competitor tests to assess if the difference between these three groups is significant. All the considered $k$-sample tests reject the null hypothesis of equality among the groups, with the only exception being the KBQD test with bootstrap when scale alternatives are used for the selection of the tuning parameter. Table S2 in section S9 of the Supplementary Material reports the obtained test statistics, critical values and/or p-values.  
\begin{table}[!ht]
\caption{Results of the considered two-sample tests when comparing the ``Adelie" and ``Chinstrap" species of the Penguin data set. For each test, the obtained test statistic, the critical value or the $p$-value, and the final answer to the null hypothesis are displayed. The KBQD tests are performed with the indicated $h$, selected via algorithm \ref{alg:h-sel}, and with \texttt{location} alternatives.}
\label{tab:peng-two-sample}
\centering
\begin{tabular}{lrrrrr}
  \hline
 Method & $h$ & Statistics & critical Value & p-value & reject $H0$ \\ 
  \hline
  Tn Sub & 0.8 & 1.346008 & 0.924375 & - & \texttt{TRUE} \\ 
  Tn Boot & 0.8 & 1.346008 & 7.30046 & - & \texttt{FALSE} \\ 
  Tn Perm & 0.4 & 0.773143 & 0.732266 & - & \texttt{TRUE} \\ 
  MMD & - & 0.012736 & 0.028894 & - & \texttt{FALSE} \\ 
  energy & - & 671.89 & - & 0.1788 & \texttt{FALSE} \\ 
  FR-WW & - & -7.705333 & - & 0.0010 & \texttt{TRUE} \\ 
  FR-KS & - & 0.963613 & - & 0.258741 & \texttt{FALSE} \\ 
  mod-KS & - & -23.15271 & - & 0.015984 & \texttt{TRUE} \\ 
   \hline
\end{tabular}
\end{table}

Considering the greater similarity between  the ``Adelie" and ``Chinstrap" species shown in Figure~\ref{fig:peng-pairs}, we want to assess the performance of the two-sample tests when comparing these two groups of observations. Considering the difference in mean in the variable ``Bill Length" between ``Adelie" and ``Chinstrap" species, we use algorithm \ref{alg:h-sel} to select $h$ with \texttt{location} alternatives. Table \ref{tab:peng-two-sample} reports the obtained test statistics, with the corresponding selected $h$ for the KBQD tests, the critical value or the $p$-value, and their answer to the null hypothesis.
The KBQD tests, with permutation and subsampling as sampling algorithm used for the computation of the critical value, together with the Friedman-Rafsky Wald-Wolfowitz test and the modified Friedman-Rafsky Kolmogorov-Smirnov test, reject the null hypothesis, while the MMD, the Energy test, KBQD tests with bootstrap and KS test do not reject the hypothesis of equality between the two distributions. Table S3 in Section S9 of the Supplementary Material report the tests statistics and critical values of the proposed $\mathrm{trace}(\hat{D}_n)$ and $T_n$ statistics obtained for the values of $h$ chosen considering all the possible alternatives in Algorithm \ref{alg:h-sel}. The performance of the KBQD test with bootstrap resampling aligns with the MMD and energy tests that likewise use bootstrap-based critical values. This pattern is consistent with the resampling literature, which notes that the finite-sample behavior of bootstrap resampling tests can be sensitive to sample-size imbalance \citep{efron1980}.

\section{Discussion and Conclusions\label{sec:conclusion}}

In this paper we present a unified framework for the study of goodness-of-fit that is based on the concept of matrix distance. Specifically, we define the kernel-based matrix distance and associated statistics for testing equality of distributions in $k$-samples. We show that the two-sample goodness-of-fit is a special case of this formulation, and propose algorithms to facilitate the computation of the test statistics. We also discuss the asymptotic distribution of the matrix distance, and thoroughly investigate the performance of the KBQD tests. The performance of the test statistics depends on a kernel tuning parameter $h$. We propose an algorithm for selecting this parameter and exemplify its use a real-data example. 

Several methods have been developed to approximate the distribution of a linear combination of chi-squared random variables (\cite{davies1980, moschopoulos1984, kume2025, coelho2020}), and several \texttt{R} packages encode methods for this purpose. For example, the packages ``\texttt{survey}" \citep{survey, lumley2004}, ``\texttt{sphunif}" \citep{sphunif, portugues2018} and ``\texttt{momentchi2}" \citep{momentchi2} all, very recently developed, contain functions encoding algorithms approximating the aforementioned distribution. 
\cite{bodenham2016} provided comparisons of different methods, including the normal approximation (see Section 6.2, p. 925 of \citet{bodenham2016}), that can be used to approximate the distribution of a linear combination of chi-squared random variables. 

In our case, our simulations indicate that the power of the tests depends on the tuning parameter $h$. Further investigation is needed to fully elucidate the role of the tuning parameter $h$ on the power of the tests. We note here that since the kernel we use is a normal probability density, we chose the covariance matrix to be $\Sigma = h^2 I$. This choice simplifies the computation, but it applies the same amount of smoothing on every component of a $d$-dimensional random vector. Alternatives include to select $\Sigma = \mathrm{diag}(h_i^2)$, $i = 1, \ldots, d$, and additional work is needed to investigate the impact of this selection on the performance characteristics of the tests, as well as the computational complexity of the algorithms.

Our simulation results indicate that the proposed tests outperform, in terms of power, the MMD and energy tests for alternatives close to the null hypothesis and for distributions with heavy tails. An example of such distributions is the family of Cauchy$(\vect{\epsilon}, I)$. The KBQD tests show an improvement in power for asymmetric alternatives, such as the skew-normal distribution, especially for small sample size and large dimension.

In summary, the proposed framework allows for goodness-of-fit testing with $k \ge 2$. Furthermore, our proposed tests perform well in terms of level and power. They are at least competitive when compared with the state-of-the-art tests in the literature, and outperform those in terms of power for contiguous alternatives, heavy tailed distributions and in higher dimensions. The proposed tests and algorithms are implemented in the software \texttt{QuadratiK} that can be used in both, \texttt{R} and \texttt{Python} environments.

\begin{appendices}

\section{Proofs}

\subsection{Proof of Proposition \ref{prop:asym-D}}
\label{app:prop3}

From Proposition \ref{prop:Dij-spectral},
\begin{equation*}
	D_n = \sum_{p=1}^\infty \lambda_p \tilde{\vect{\phi}}_{n,p} \tilde{\vect{\phi}}_{n,p}^\top ,
\end{equation*}
where $\tilde{\vect{\phi}}_{n,p}$ has entries given by    
\begin{equation*}
	\bar{\phi}_{n,p,i} = \frac{1}{n_i}\sum_{j=1}^{n_i} \phi_p(\vect{x}^{(i)}_j).
\end{equation*}
Note that
\begin{equation*}
	\mathbb{E}(\bar{\phi}_{n,p,i}) =  \frac{1}{n_i}\sum_{j=1}^{n_i} \mathbb{E}(\phi_p(\vect{x}^{(i)}_j)) = 0,
\end{equation*}
\begin{equation*}
	\mathrm{Var}(\bar{\phi}_{n,p,i}) = \frac{1}{n_i^2} \left[ \sum_{j=1}^{n_i} \mathbb{E}(\phi_p^2(\vect{x}^{(i)}_j)) +  \sum_{j=1}^{n_i} \sum_{\ell\not = j=1}^{n_i} \mathbb{E}(\phi_p(\vect{x}^{(i)}_j)\phi_p(\vect{x}^{(i)}_\ell))\right]= \frac{1}{n_i},
\end{equation*}
and for $1 \le i \not= j \le k$
\begin{align*}
	\mathrm{Cov}(\bar{\phi}_{n,p,i}, \bar{\phi}_{n,p,j}) = & \mathbb{E} \left[ \left( \frac{1}{n_i}\sum_{\ell=1}^{n_i} \phi_p(\vect{x}^{(i)}_\ell)\right)\left(\frac{1}{n_j}\sum_{r=1}^{n_j} \phi_p(\vect{x}^{(j)}_r)\right) \right] \\
 = &  \frac{1}{n_i n_j}\sum_{\ell=1}^{n_i} \sum_{r=1}^{n_j} \mathbb{E}(\phi_p(\vect{x}^{(i)}_\ell)\phi_p(\vect{x}^{(j)}_r)) = 0.
\end{align*}
Let $\rho_i = \lim_{n,n_i \to \infty} n_i/n$, with $0 < \rho_i < 1$, and let $\mat{V} = \mathrm{diag}(1/\rho_1,\ldots,1/\rho_k)$.
Assumptions (i)-(ii) ensures finite second moments, then by the multivariate Central Limit Theorem we have that 
\begin{equation*}
    \sqrt{n} \tilde{\vect{\phi}}_{n,p} \stackrel{d}{\longrightarrow} N_k(\vect{0},\mat{V}). 
\end{equation*}
Let
\begin{equation*}
\mat{D}_n^{(m)}=\sum_{p=1}^{m}\lambda_p\,\tilde{\vect{\phi}}_{n,p}\tilde{\vect{\phi}}_{n,p}^{\top},
\qquad
\mat{R}_{n,m}=\sum_{p>m}\lambda_p\,\tilde{\vect{\phi}}_{n,p}\tilde{\vect{\phi}}_{n,p}^{\top},
\end{equation*}
so that ${\mat{D}}_n={\mat{D}}_n^{(m)}+\mat{R}_{n,m}$.
For a finite number of indices $p\le m$, by the continuous mapping theorem
\begin{equation*}
    n\,\hat{\mat{D}}_n^{(m)}
=\sum_{p\le m}\lambda_p\bigl(\sqrt n\,\tilde{\vect{\phi}}_{n,p}\bigr)\bigl(\sqrt n\,\tilde{\vect{\phi}}_{n,p}\bigr)^{\!\top}
\ \stackrel{d}{\longrightarrow}\
\sum_{p\le m}\lambda_p\,W_{k,p},
\end{equation*}
with independent random variables following a Wishart distribution with covariance
matrix $V$ and $1$ degree of freedom,  $ W_{k,p}\sim\mathcal W_k(1,\mathbf V)$.

\noindent Under assumptions (i)–(ii), we have that
\begin{equation*}
\sum_{p=1}^{\infty}\lambda_p
=\int K_{\bar F}(\mathbf t,\mathbf t)\,d\bar F(\mathbf t)<\infty,
\qquad
\sum_{p=1}^{\infty}\lambda_p^2<\infty.
\end{equation*}
Then, the remainder $\mat{R}_{n,m}$ is positive semidefinite, the operator norm is such that $\|\mat{R}_{n,m}\|_{\mathrm{op}}\le \operatorname{tr}(\mat{R}_{n,m})$ and
\begin{equation*}
\mathbb E\!\left[\operatorname{tr}\!\bigl(n\,{R}_{n,m}\bigr)\right]
= \sum_{p>m}\lambda_p\ \mathbb E\!\left[n\,\|\tilde{\vect{\phi}}_{n,p}\|^2\right]
= \Bigl(\sum_{i=1}^k \frac{n}{n_i}\Bigr)\ \sum_{p>m}\lambda_p.
\end{equation*}
Since $n_i/n\to\rho_i\in(0,1)$, there exists a constant $C>0$ and $N$ such that
$\sum_{i=1}^k n/n_i\le C$ for all $n\ge N$.
Thus, for all $n\ge N$,
\begin{equation*}
\mathbb E\!\left[\operatorname{tr}\!\bigl(n\,\mat{R}_{n,m}\bigr)\right]
\le C\,\sum_{p>m}\lambda_p.
\end{equation*}
By Markov’s inequality,
\begin{equation*}
\sup_{n\ge N}\ \mathbb P\!\left(\bigl\|n\,\mat{R}_{n,m}\bigr\|_{\mathrm{op}}>\varepsilon\right)
\ \le\ \frac{C}{\varepsilon}\,\sum_{p>m}\lambda_p
\ \xrightarrow[m\to\infty]{}\ 0,
\end{equation*}
so $n\,\mat{R}_{n,m}\to 0$ in probability, uniformly in $n$, as $m\to\infty$.

\noindent Combining the results obtained and applying Slutsky's theorem, we have that
\begin{equation*}
n{\mat{D}}_n
= n{\mat{D}}_n^{(m)} + n\,\mat{R}_{n,m}
\ \stackrel{d}{\longrightarrow}\
\sum_{p\le m}\lambda_p\,W_{k,p}
\end{equation*}
and letting $m\to\infty$, we obtain
\begin{equation*}
n{\mat{D}}_n \ \stackrel{d}{\longrightarrow}\
\sum_{p=1}^{\infty}\lambda_p\,W_{k,p} \,=:\, W^{\ast}.
\end{equation*}
Because each $\lambda_p W_{k,p}$ is positive semidefinite and independent
\begin{equation*}
\mathbb E\!\left[\operatorname{tr}\!\left(\sum_{p=1}^{\infty}\lambda_p\mat{W}_{k,p}\right)\right]
= \operatorname{tr}(\mat{V})\,\sum_{p=1}^{\infty}\lambda_p \;<\;\infty,
\end{equation*}
hence the series $\sum_{p=1}^{\infty}\lambda_p\mat{W}_{k,p}$
converges almost surely in trace norm (monotone convergence with finite expectation).
Finally, this implies that 
\begin{equation*}
    n{\mat{D}}_n = \sum_{p=1}^\infty \lambda_p n  \tilde{\vect{\phi}}_{n,p}  \tilde{\vect{\phi}}_{n,p}^\top \stackrel{d}{\longrightarrow} \sum_{p=1}^\infty \lambda_p W_{k,p}
\end{equation*}
where $W_p \sim \mathcal{W}_{k,p}(1, \mat{V})$ are independent copies of the Wishart distribution of dimension $k$, one degree of freedom and covariance matrix $\mat{V}$.

\subsection{Proof of Proposition \ref{prop:asym-trace}}
\label{app:asym-trace}

We can write the statistic $S_n = \mathrm{trace(\hat{D}_n)}$ as follows: 
\begin{equation*}
    S_n = \sum_{i=1}^k \hat{D}_{ii},
\end{equation*}
\begin{equation*}
	\hat{D}_{ii} = \frac{1}{n_i (n_i -1)} \sum_{\ell=1}^{n_i}\sum_{r\not= \ell}^{n_i}K_{\bar{F}}(\vect{x}^{(i)}_\ell,\vect{x}^{(i)}_r), \qquad \mbox{ }i = j.
\end{equation*}
The statistic $\hat{D}_{ii}$ is a $U$-statistic, and its limiting distribution is given as $\sum_{p=1}^\infty \lambda_p (Z_{ip}^2 -1)$ (see \cite{Lindsay2014}, p-397; complete proof is given in \cite{serfling1980}, Section 5.5.2, pp 195-199, and in \cite{Liu1995}). 
Because $\rho_i = \lim_{n_i, n \to \infty} n_i/n$ and $1/n_i(n_i - 1) \approx 1/n_i n_i$ for $n_i \to \infty$, we obtain
\begin{align*}
    n\hat{D}_{ii} = &  \frac{n}{n_i (n_i - 1)} \sum_{\ell=1}^{n_i} \sum_{r \not = \ell}^{n_i} K_{\bar{F}}(\vect{x}^{(i)}_\ell,\vect{x}^{(i)}_r) \\ 
    & \stackrel{d}{\longrightarrow}  \sum_{p=1}^{\infty} \lambda_p \left(\frac{1}{\rho_i} Z_{ip}^2 - \frac{1}{\rho_i}\right), 
\end{align*}
where $Z_{ip}$, $i \in \{1,\ldots k\}$, are independent identically distributed $N(0,1)$ random variables. This convergence is guaranteed under conditions $(i), (ii)$, as well as because of the kernel centering  and orthogonality of eigenvectors.

\noindent Additionally, $D_{ii}, D_{jj}$, $i \not= j$, $i,j \in \{1, \ldots, k\}$, are independent random variables (the $k$-samples are independent). Therefore, using Slutsky's theorem
\begin{equation*}
    S_n \stackrel{d}{\longrightarrow} \sum_{p=1}^\infty \lambda_p \left( \sum_{i=1}^k \left(\frac{1}{\rho_i} Z_{ip}^2 - \frac{1}{\rho_i}\right)\right), \quad \mbox{ as } n_i, n \to \infty.
\end{equation*}

\subsection{Proof of Proposition \ref{prop:asym-norm-test}}
\label{app:prop5}

Consider that, under the null hypothesis, all samples from $F_i$, $i=1, \ldots, k$, are independent and identically distributed from $\bar{F}$, and samples are mutually independent. Considering that the kernel is centered with respect to $\bar{F}$, we have that
\begin{equation*}
    \int K_{\bar{F}}(\vect{s},\vect{t}) d\bar{F}(\vect{t}) = \int K_{\bar{F}}(\vect{s},\vect{t}) dF_i(\vect{t}) = 0.
\end{equation*}
This implies that the centered kernel $\bar{F}$ has the same eigenvalues and eigenfunctions with respect to each $F_i$ under the null hypothesis. Considering assumptions (i), (ii) and (iii), by Mercer's theorem we can write the spectral decomposition in $L_2(\bar{F})$ of the kernel $K_{\bar{F}}(\vect{x}, \vect{y}) = \sum_{p=1}^\infty \lambda_p \phi_p(\vect{x})\phi_p(\vect{y})$ with $\int \phi_p(\vect{x})\phi_q(\vect{x})d\bar{F}(\vect{x}) = \delta_{pq}$ and $\lambda_p \ge 0$. Then, we can rewrite $nT_{n}$ as
\begin{align*}
    nT_{n} =  & \sum_{p=1}^\infty \lambda_p  \left\{ (k-1)\sum_{i=1}^k \frac{n}{n_i(n_i-1)} \sum_{\ell=1}^{n_i} \sum_{r \not=\ell}^{n_i} \phi_p(\vect{x}_\ell^{(i)})\phi_p(\vect{x}_r^{(i)}) \right. \\
    & \left.- 2 \sum_{i=1}^k \sum_{j >i}^k \frac{n}{n_i n_j} \sum_{\ell=1}^{n_i} \sum_{r=1}^{n_j} \phi_p(\vect{x}_\ell^{(i)})\phi_p(\vect{x}_r^{(j
    )})\right\}.
\end{align*}
Notice that, for $i,j \in \{1,\ldots,k\}$
\begin{align*}
     \sum_{\ell=1}^{n_i} \sum_{r \not=\ell}^{n_i} \phi_p(\vect{x}_\ell^{(i)})\phi_p(\vect{x}_r^{(i)}) & = \left( \sum_{\ell=1}^{n_i} \phi_p(\vect{x}_\ell^{(i)}) \right)^2 - \sum_{\ell=1}^{n_i} \phi_p^2(\vect{x}_\ell^{(i)}), \\
     \sum_{\ell=1}^{n_i} \sum_{r=1}^{n_j} \phi_p(\vect{x}_\ell^{(i)})\phi_p(\vect{x}_r^{(j)}) & = \left( \sum_{\ell=1}^{n_i} \phi_p(\vect{x}_\ell^{(i)}) \right) \left( \sum_{r=1}^{n_j} \phi_p(\vect{x}_r^{(j)}) \right).
\end{align*}
Considering the two equalities above, as well as that $\frac{n}{n_i(n_i-1)}$ is equivalent to $n/n_i^2$ for large $n_i$ and $n$, we can further rewrite $nT_{n}$ as
\begin{align*}
    nT_{n} =  &\sum_{p=1}^\infty \lambda_p \left\{ (k-1)\sum_{i=1}^k \frac{1}{\rho_i} \left[\left(\frac{1}{\sqrt{n_i}} \sum_{\ell=1}^{n_i} \phi_p(\vect{x}_\ell^{(i)}) \right)^2 - \frac{1}{n_i}\sum_{\ell=1}^{n_i} \phi_p^2(\vect{x}_\ell^{(i)})\right] \right.\\
    & \left. -2 \sum_{i=1}^k \sum_{j>i}^k \frac{1}{\sqrt{\rho_i\rho_j}} \left( \frac{1}{\sqrt{n_i}}\sum_{\ell=1}^{n_i} \phi_p(\vect{x}_\ell^{(i)}) \right) \left( \frac{1}{\sqrt{n_j}}\sum_{r=1}^{n_j} \phi_p(\vect{x}_r^{(j)}) \right) \right\}\\
    = & \sum_{p=1}^\infty \lambda_p \left\{ \sum_{i=1}^k \sum_{j>i}^k  \left(\frac{1}{\sqrt{\rho_i}} \frac{1}{\sqrt{n_i}} \sum_{\ell=1}^{n_i} \phi_p(\vect{x}_\ell^{(i)}) - \frac{1}{\sqrt{\rho_j}} \frac{1}{\sqrt{n_j}}\sum_{r=1}^{n_j} \phi_p(\vect{x}_r^{(j)})\right)^2 \right.\\
    & - \left. \sum_{i=1}^k \frac{1}{\rho_i} \frac{1}{n_i} \sum_{\ell=1}^{n_i} \phi_p^2(\vect{x}_\ell^{(i)}) \right\} .
\end{align*}
Notice that here the following identity has been used 
$$(k-1)\sum_i a_i^2 - 2 \sum_{i<j} a_ia_j = \sum_{i<j} (a_i - a_j)^2 \mbox{ with } a_i = \rho^{-1/2} n^{-1/2} \sum_{\ell=1}^{n_i} \phi_p(\vect{x}_{\ell}^{(i)}).$$
Recall that $\mathbb{E}[\phi_p(\vect{x})] = 0$ and $\mathbb{E}[\phi_p(\vect{x})\phi_q(\vect{x})] = \delta_{pq}$, then employing the weak law of large numbers and the Lindeberg-Levy Central Limit Theorem, we have that
\begin{align*}
     &\frac{1}{n_i} \sum_{\ell=1}^{n_i} \phi_p^2(\vect{x}_\ell^{(i)}) \stackrel{p}{\longrightarrow} 1 \mbox{ as } n_i\to \infty, \\
     &\frac{1}{\sqrt{n_i}} \sum_{\ell=1}^{n_i} \phi_p(\vect{x}_\ell^{(i)}) \stackrel{d}{\longrightarrow} Z_{ip} \sim N(0,1) \mbox{ as } n\to \infty.
\end{align*}
Furthermore, since the $k$ groups $(\vect{x}_1^{(1)}, \ldots, \vect{x}_{n_1}^{(1)}), \ldots, (\vect{x}_1^{(k)}, \ldots, \vect{x}_{n_k}^{(k)})$ are mutually independent we have for each $1 \le i \not=j \le k$
\begin{equation*}
\mathrm{Cov}\bigg(\frac{1}{\sqrt{n_i}}  \sum_{\ell=1}^{n_i} \phi_p(\vect{x}_\ell^{(i)}), \frac{1}{\sqrt{n_j}} \sum_{r=1}^{n_j} \phi_p(\vect{y}_r^{(j)}) \bigg) = 0.   
\end{equation*}
This implies that the limit components $(Z_{1p}, \ldots, Z_{kp})$ are independent. 

\noindent
Let $P \in \mathbb{N}$. We define a truncated version of the statistic
\begin{align*}
    nT_{n,P} = & \sum_{p=1}^P \lambda_p \left\{ \sum_{i=1}^k \sum_{j>i}^k  \left(\frac{1}{\sqrt{\rho_i}} \frac{1}{\sqrt{n_i}} \sum_{\ell=1}^{n_i} \phi_p(\vect{x}_\ell^{(i)}) - \frac{1}{\sqrt{\rho_j}} \frac{1}{\sqrt{n_j}}\sum_{r=1}^{n_j} \phi_p(\vect{x}_r^{(j)})\right)^2 \right.\\
    & - \left. \sum_{i=1}^k \frac{1}{\rho_i} \frac{1}{n_i} \sum_{\ell=1}^{n_i} \phi_p^2(\vect{x}_\ell^{(i)}) \right\} .
\end{align*}
We know that $\left(\frac{1}{\sqrt{n_i}} \sum_{\ell=1}^{n_i} \phi_p(\vect{x}_\ell^{(i)}); \frac{1}{n_i} \sum_{\ell=1}^{n_i} \phi_p^2(\vect{x}_\ell^{(i)}) \right)$ converges jointly in distribution to $(Z_{ip}, 1)$. 
Hence, by the continuous mapping theorem and Slutsky's theorem, we have that
\begin{equation*}
    nT_{n,P}  \stackrel{d}{\longrightarrow} T_P :=  \sum_{p=1}^P \lambda_p \left[ \sum_{i=1}^k \sum_{j>i}^k \left( \frac{1}{\sqrt{\rho_i}}Z_{ip}- \frac{1}{\sqrt{\rho_j}}Z_{jp}\right)^2 - \sum_{i=1}^k \frac{1}{\rho_i} \right] \mbox{ as } n_i,n \to \infty.
\end{equation*}
For each fixed $P$, the terms in the sum over $p$ are independent, then $T_P$ is a finite sum of independent centered quadratic forms of independent Gaussian random variables. 

\noindent Let's now consider the tail portion
$$K^P(\vect{x}, \vect{y}) = \sum_{p>P} \lambda_p \phi_p(\vect{x}) \phi_p(\vect{y}),$$
and let $R_{n,P} = nT_n - nT_{n,P}$.
By the result in \citet[Section 5.5]{serfling1980}, there exist a constant $C = C(k, \rho_1, \ldots, \rho_k)$ independent of $n$ and $P$ such that 
\begin{align*}
\sup_{n \ge 1} \mathrm{Var}(R_{n,P}) \le & C \| K^P(\vect{x}, \vect{y}) \|^2_{L_2(\bar{F})} \\
= & C \sum_{p>P} \lambda_p^2 < \infty.
\end{align*}
Since $\sum_{p \ge 1} \lambda_p^2 < \infty$ by assumption (ii), by Chebyshev inequality 
\begin{equation*}
    \lim_{p \to \infty} \sup_{n \ge 1} \mathbb{P} \left(|R_{n,P}|>\varepsilon\right) = 0, \qquad \forall \varepsilon>0.
\end{equation*}
That is, $R_{n,P} \stackrel{p}{\longrightarrow} 0$ as $p \to \infty$, uniformly for large $n$. Then, for fixed $P$, $nT_n = nT_{n,P} + R_{n,P} \stackrel{n \to \infty}{\longrightarrow} T_P$. 

\noindent
Finally, as $P \to \infty$ the finite sum $T_P$ converges in $L_2$, and hence in distribution, to
\begin{equation*}
    \chi^\ast = \sum_{p=1}^\infty \lambda_p \left[ \sum_{i=1}^k \sum_{j>i}^k \left( \frac{1}{\sqrt{\rho_i}}Z_{ip}- \frac{1}{\sqrt{\rho_j}}Z_{jp}\right)^2 - \sum_{i=1}^k \frac{1}{\rho_i} \right],
\end{equation*}
using the uniform tail bound.

\subsection{Theorem 3.1 \cite[Annals of Statistics, p.988]{Lindsay2008}}
\label{app:sec:therem3.1-lindsay}

A nonnegative definite kernel, that satisfies the condition 
\begin{equation*}
\int_{\mathcal{S}} \int_{\mathcal{S}}   K^2(\vect{x},\vect{y}) dM(\vect{x})dM(\vect{y}) < \infty. 
\end{equation*}
can be written as 
\begin{equation*}
    K(\vect{x},\vect{y}) = \sum_{j=1}^\infty \lambda_j \phi_j(\vect{x})\phi_j(\vect{y}),
\end{equation*}
where $\lambda_j$ are the eigenvalues and $\phi_j$ are the corresponding normalized eigenvectors of $K$ under the baseline measure $M$. The series above converge strongly to $K$; that is, for every $g$ in $L_2$ 
\begin{equation*}
    \lim_{n\to \infty} \int_{\mathcal{S}} \left( \int_{\mathcal{S}}   K(\vect{x},\vect{y}) g(\vect{y}) dM(\vect{y}) - \sum_{j=1}^n \int_{\mathcal{S}} \lambda_j \phi_j(\vect{x})\phi_j(\vect{y})g(\vect{y})dM(\vect{y}) \right)^2 dM(\vect{x}) = 0
\end{equation*}

\end{appendices}

\section*{Supplementary information}

This manuscript is accompanied by a Supplementary Material document, which includes the derivation of the two-sample statistic as a special case, and provides further proofs and simulation results that complement the findings presented in the main manuscript.

\section*{Acknowledgements}
The first author would like to acknowledge KALEIDA Health Foundation award 82114, that supported the work of the second author.

\bibliography{kernel_test}

\includepdf[pages=-,pagecommand=\thispagestyle{empty},fitpaper=true]{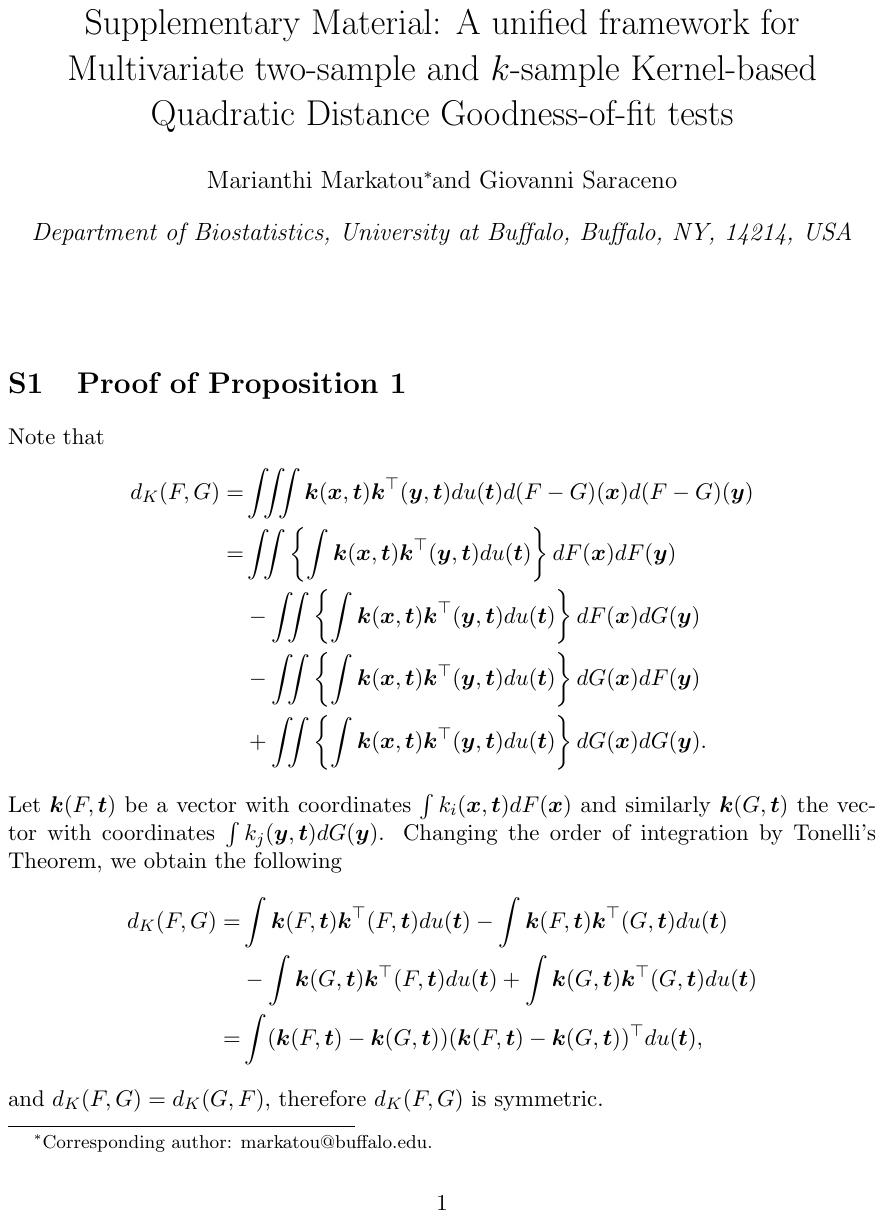}

\end{document}